\documentclass[journal,9pt]{IEEEtran}
\usepackage{amsmath,amsfonts}
\usepackage{algorithmic}
\usepackage{algorithm}
\usepackage{array}
\usepackage{microtype}
\usepackage[caption=false,font=normalsize,labelfont=sf,textfont=sf]{subfig}
\usepackage{booktabs}
\usepackage{textcomp}
\usepackage{stfloats}
\usepackage{url}
\usepackage{verbatim}
\usepackage{graphicx}
\usepackage{cite}
\usepackage{makecell}
\usepackage{hyperref}
\begin{document}

\title{\huge{Complex Permittivity Extraction of Polymer Materials Using Gradient-Enhanced NSGA-II Algorithm}}

\author{Hanqing Zhang, Zhuowei Li,~\IEEEmembership{Member,~IEEE,} Jiliang Zhang,~\IEEEmembership{Senior Member,~IEEE,} Xi Liao,~\IEEEmembership{Senior Member,~IEEE,} Yang Wang~\IEEEmembership{Senior Member,~IEEE}

\thanks{Corresponding Author: Jiliang Zhang (zhangjiliang1@mail.neu.edu.cn)}%
\thanks{H. Zhang, Z. Li and J. Zhang are with the College of Information Science and Engineering, Northeastern University, China, 110819.}%
\thanks{X. Liao is with the School of Communication and Information Engineering, Chongqing University of Posts and Telecommunications, Chongqing, China, 400065.}%
\thanks{Y. Wang is with the International College, Chongqing University of Posts and Telecommunications, Chongqing, China, 400065}%

\thanks{This work was supported by the National Key R\&D Program of China under Grant 2025YFE0122200, the National Natural Science Foundation of China (NSFC) under Grant 62573096 and 62401644, the Postdoctoral Fellowship Program and China Postdoctoral Science Foundation under Grant BX20250343, the LiaoNing Revitalization Talents Program under Grant XLYC2403116, the Opening Fund of Liaoning Key Laboratory of Urban and Architectural Digital Technology under Grant UADT2024A05, the Fundamental Research Funds for the Central Universities under Grant N25XQD031 and N25ZLH010, and the Basic Scientific Research Fund of The State Key Laboratory of Synthetical Automation for Process Industries.}}



\maketitle
\begin{abstract}
This paper presents gradient-enhanced non-dominated sorting genetic algorithm II (G-NSGA-II) to address the challenges of local optima and solution non-uniqueness in the complex permittivity extraction problem for the first time. This adaptive hybrid algorithm integrates the global exploration capability of NSGA-II with gradient-based local refinement, triggered by a population-stagnation detection mechanism. Furthermore, multi-dimensional constraints are incorporated by jointly optimizing transmission and reflection coefficients across multiple sample thicknesses. Experimental validation conducted on six typical polymers in the 20--40 GHz band demonstrates that the retrieved relative permittivity and thicknesses are in high agreement with literature values and physical measurements. Compared to standard heuristic and gradient-based algorithms, the proposed G-NSGA-II reduces the number of generations required for convergence by approximately 50\%. This significant improvement in speed, combined with enhanced robustness, provides a highly reliable and efficient solution for broadband dielectric characterization in architectural and electromagnetic engineering. The simple measurement method and the proposed efficient algorithm allow for a rapid evalutaion of wireless performance within indoor environments. This approach serves as a valuable tool for optimizing existing wireless layouts and improving network performance.
\end{abstract}


\section{Introduction}
\IEEEPARstart{T}{he} integration of 5G and 6G millimeter-wave systems into indoor environments has intensified interest in the electromagnetic properties of building materials \cite{populor}. As internal structures and infill patterns of polymers significantly influence their effective permittivity, accurate characterization of material-wave interaction is vital for predicting transmission, reflection, and attenuation \cite{bwp1, Persad_Infill_2024}. Accurate electromagnetic parameters are fundamental to metrics such as spatially averaged capacity (SAC) \cite{bwp2}, enabling precise channel simulation \cite{Li_SiC} and interference prediction, such as far-end crosstalk \cite{Sanchez_Crosstalk_2026}. Consequently, achieving high-precision, non-destructive extraction of complex permittivity and permeability across a wide spectrum remains a primary challenge\cite{Alhassoon_Extract}.

Traditional characterization relies on waveguide and resonant cavity techniques \cite{ref9, ref10, ref11, ref12, ref13, ref14}, which, despite their precision in millimeter-wave bands, require destructive and rigorous sample machining to avoid air-gap errors \cite{ref9, ref13}. In contrast, the free-space method is more suitable for the evaluation of diverse building materials and composites due to its operational simplicity and non-destructive nature, which eliminates the need for precise sample fabrication \cite{Chang_FreeSpace}. To further enhance this method and simplify the measurement setup, various approaches have been proposed, including time-domain gating, analytic solutions, and novel rotation-based or standardless calibration techniques \cite{Brandl_Rotation, ref15, ref16}. However, these methods often suffer from convergence instability, sensitivity to initial guesses, or limited validity ranges \cite{ref17, ref18, ref19, ref20}.

A major hurdle in free-space characterization is the inherent ill-posedness of inverse problems. Standard iterative solvers frequently stall in local optima, while global heuristics like genetic algorithms (GA) and particle swarm optimization (PSO) exhibit slow late-stage convergence \cite{ref21, ref22, ref23}. To ensure mathematical uniqueness and physical reliability, a hybrid strategy balancing global exploration with local refinement is necessary. Furthermore, incorporating multi-dimensional constraints—such as jointly optimizing transmission and reflection coefficients—is essential for robust broadband extraction.

To address these challenges, this paper performs electromagnetic parameter extraction based on free-space method measurements, utilizing a gradient-enhanced non-dominated sorting genetic algorithm II (G-NSGA-II). The main contributions and novelties of our work are summarized as follows:
\begin{itemize}
    \item Using the free-space method, we measured the transmission and reflection coefficients of six representative polymers: polyamide (PA), polycarbonate (PC), polyethylene (PE), polyethylene terephthalate (PET), polytetrafluoroethylene (PTFE), and polypropylene (PP).
    \item G-NSGA-II was applied to tackle the challenge of multiple local optima encountered in electromagnetic parameter extraction.
    \item The convergence speed of the proposed G-NSGA-II was compared with standard GA and NSGA-II algorithms, demonstrating the high efficiency of the designed extraction algorithm.
\end{itemize}

Although this study focuses on representative polymers, the proposed algorithmic framework offers a scalable and efficient solution for extracting the electromagnetic properties of various building materials. This capability is essential for systematically populating material databases for existing indoor environments. Such comprehensive material characterization is a prerequisite for high-fidelity channel modeling, ray-tracing simulations, and the construction of electromagnetic digital twins. Consequently, the precise permittivity extraction method provided in this work serves as a vital tool for engineers to predict signal interaction with indoor obstacles, thereby facilitating the optimized deployment of next-generation wireless systems.

The remainder of this paper is organized as follows. Section II details the construction of the measurement platform and the theoretical model. Section III elaborates on the specific implementation of the G-NSGA-II algorithm. Section IV presents a comparative analysis of the complex permittivity retrieved by the proposed method against literature values, along with a convergence speed comparison between the proposed algorithm and existing counterparts. Conclusions are drawn in Section V.

\section{Measurement}
\subsection{Measurement Setup}
\begin{figure}[!t]
\centering
\includegraphics[width=0.48\textwidth]{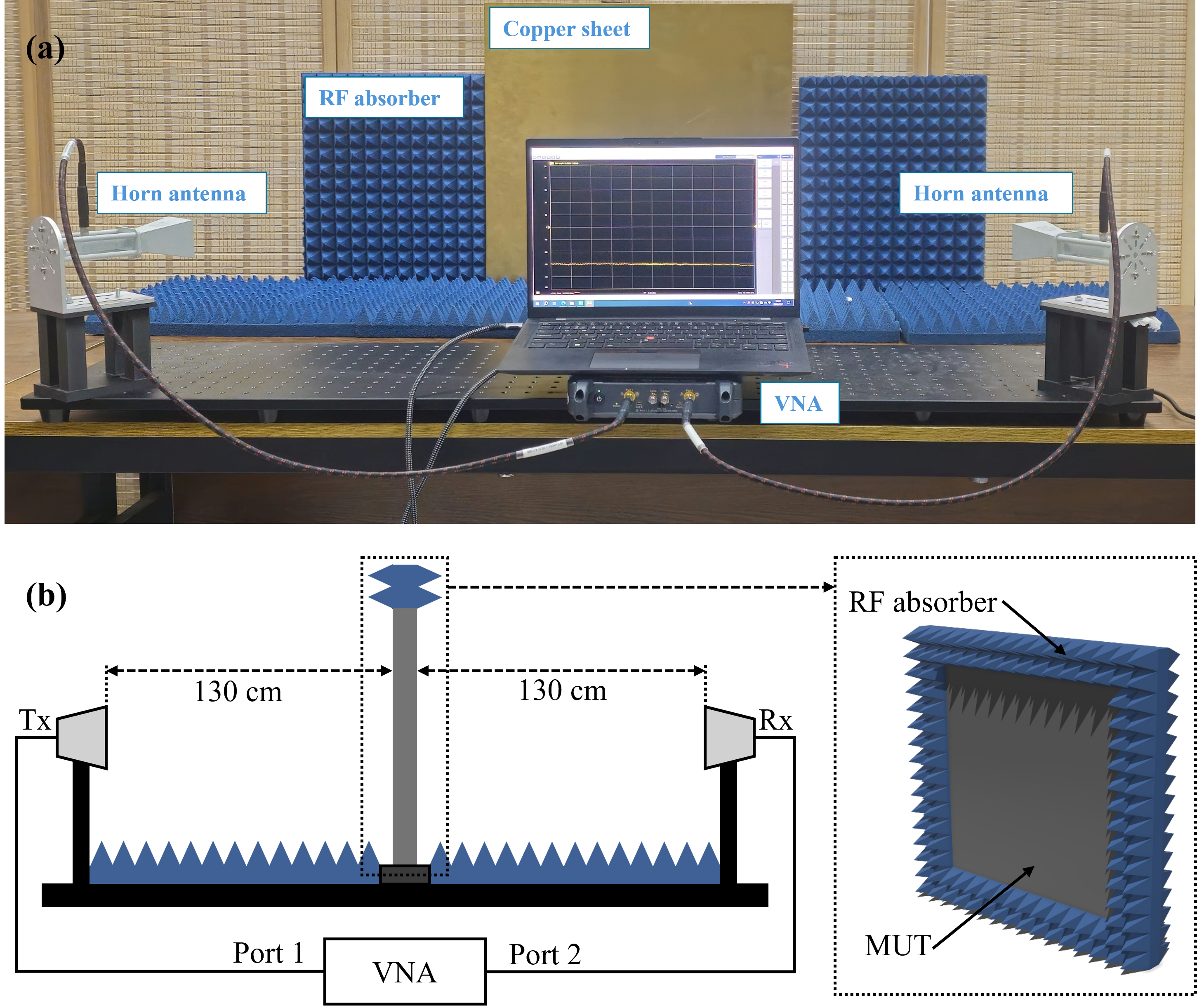}
\caption{(a) Photograph of the experimental setup using a metal sheet as a total reflection reference. (b) Schematic diagram of the transmission measurement system for the polymer slab.}
\label{fig_1}
\end{figure}
A complex permittivity measurement system based on the free-space method was established. The setup utilizes a Keysight P9377B vector network analyzer (VNA) managed by host software, with broadband linearly polarized horn antennas operating in the 20--40 GHz band. VNA configuration parameters are summarized in Table~\ref{measurement parameters} to balance efficiency and dynamic range. Before measurement, systematic errors from cables and connectors were eliminated via a full-port calibration using a Keysight N4693D electronic module.

Fig.~\ref{fig_1} illustrates the measurement platform, which was secured on an optical table using custom 3D-printed fixtures for precise alignment. In the transmission setup, antennas were aligned along the line-of-sight with the sample ($300\times300\text{ mm}$ slab) positioned at the midpoint; the reference S-parameter was obtained with the sample removed. For reflection, antennas were arranged symmetrically on a semicircular arc relative to the sample's surface normal, using a metal sheet to establish the total-reflection reference.

To satisfy far-field conditions, the antenna-to-sample distance was maintained above 130 cm. RF absorbers were strategically placed on the table surface, antenna back-lobes, and sample periphery to suppress multipath reflections and environmental noise.

Six low-loss polymers (PA, PC, PE, PET, PTFE, and PP) were selected due to their prevalence in indoor environments and potential as carrier media for integrated antenna arrays \cite{material_antenna}. For each material, two samples of different thicknesses were prepared. This approach introduces multi-dimensional constraints into the inverse algorithm, enhancing the robustness of the electromagnetic parameter extraction.

\subsection{Wave Propagation Model}
As the investigated materials are non-ionizing and non-magnetic, we assume zero free charge density and free-space magnetic permeability. Consequently, the electromagnetic properties are characterized by the complex relative permittivity $\epsilon_r$ \cite{ref24}

\begin{equation}
\label{definition of complex permittivity}
\epsilon_r=\epsilon-\text{i}\frac{\sigma}{\epsilon_0\omega},
\end{equation}
where $\epsilon$ is the relative permittivity, $\epsilon_0$ is the free-space permittivity, and $\omega$ is the angular frequency. The frequency-dependent conductivity $\sigma$ is modeled as

\begin{equation}
\label{definition of complex conductivity}
\sigma=cf_{\text{GHz}}^d,
\end{equation}
where $c$ and $d$ are material-specific constants and $f_{\text{GHz}}$ is the frequency in GHz.

For waves incident from air onto the material, the Fresnel reflection coefficients for TE and TM polarizations are

\begin{equation}
\label{definition of rTE}
r_{\text{TE}}=\frac{\cos{\theta}-\sqrt{\epsilon_r-\sin^2{\theta}}}{\cos{\theta}+\sqrt{\epsilon_r-\sin^2{\theta}}},
\end{equation}

\begin{equation}
\label{definition of rTM}
r_{\text{TM}}=\frac{\epsilon_r\cos{\theta}-\sqrt{\epsilon_r-\sin^2{\theta}}}{\epsilon_r\cos{\theta}+\sqrt{\epsilon_r-\sin^2{\theta}}},
\end{equation}

\begin{table}[!t]
\caption{Parameters Of Measurement System\label{measurement parameters}}
\centering
\begin{tabular}{cc} 
\toprule  
\textbf{Indicator parameters} & \textbf{Parameter value}\\
\midrule  
Center frequency & 30 GHz\\
Start frequency & 20 GHz\\
Stop frequency & 40 GHz\\
Number of sweep points & 1601\\
Frequency resolution & 12.5 MHz\\
Transmitting power & 0 dBm\\
Antenna height & 160 mm\\
\bottomrule 
\end{tabular}
\end{table}
where $\theta$ is the incidence angle. For a single-layer slab of thickness $h$, the overall reflection and transmission coefficients are modeled as \cite{ref24}

\begin{equation}
\label{definition of R}
R=\frac{r(1-e^{-i2q})}{1-r^2e^{-2iq}},
\end{equation}

\begin{equation}
\label{definition of T}
T=\frac{(1-r)(1-e^{-iq})}{1-r^2e^{-2iq}},
\end{equation}

\begin{equation}
\label{definition of q}
q=\frac{2\pi h}{\lambda}\sqrt{\epsilon_r-\sin^2{\theta}},
\end{equation}
where $r$ is the polarization-dependent Fresnel coefficient.

Using VNA-measured S-parameters, the slab's reflection and transmission coefficients are determined per \cite{ref15}

\begin{equation}
\label{definition of Rmeasure}
R_{\text{measure}}=\frac{S_{21}^{\text{MUT}}}{S_{21}^{\text{air}}},
\end{equation}

\begin{equation}
\label{definition of Tmeasure}
T_{\text{measure}}=\frac{S_{21}^\text{{MUT}}}{S_{21}^{\text{metal}}},
\end{equation}
where $S^{\text{MUT}}_{21}$, $S_{21}^{\text{air}}$, and $S_{21}^{\text{metal}}$ denote measurements for the material under test (MUT), air, and a metal plate reference, respectively.

\begin{figure}[!t]
\centering
\includegraphics[width=0.4\textwidth]{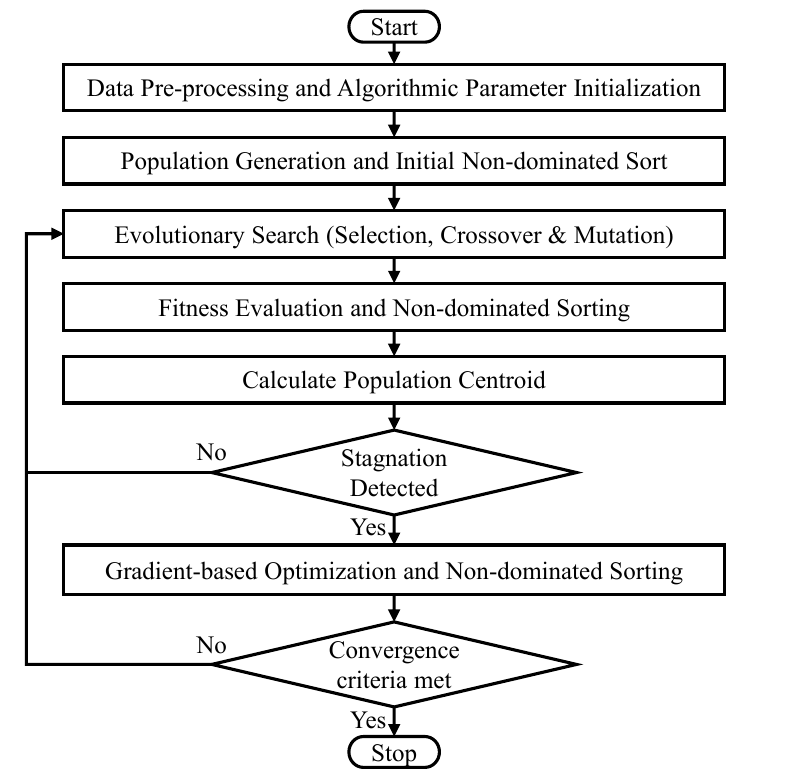}
\caption{Flowchart of the proposed G-NSGA-II algorithm.}
\label{fig:flowchart}
\end{figure}

Measured data are processed via time-gating \cite{ref15}. Frequency-domain data are converted to the time domain using the inverse fast Fourier transform to isolate and eliminate environmental multipath interference. To suppress edge effects from time-gating and Fourier transforms, 80 data points are discarded from both frequency boundaries, ensuring result integrity.

\section{Optimization Algorithm}
The extraction of $\epsilon_r$ and $h$ using measured transmission coefficients, reflection coefficients, and theoretical models constitutes a typical nonlinear and non-convex optimization problem. Given the extensive use of heuristic algorithms in this domain, this study employs an improved NSGA-II with an elitist strategy.

NSGA-II is a multi-objective evolutionary algorithm based on Pareto optimality \cite{ref25}. The algorithm evaluates the quality of individuals using fast non-dominated sorting, where all non-dominated solutions constitute the first Pareto front. To ensure population diversity, crowding distance is employed to regulate the density of solutions on each Pareto front. This approach effectively avoids common issues in the extraction process, such as multi-valued solutions and local optima traps, thereby guaranteeing the reliability of the results. 

Although the NSGA-II algorithm excels in global search, it often struggles to precisely identify the global optimum when its neighborhood contains multiple suboptimal solutions with similar objective function values. Consequently, the algorithm is prone to stagnation, oscillation, or premature convergence near these suboptimal points \cite{ref26}. In the context of electromagnetic parameter extraction, the material's transmission and reflection coefficients are jointly determined by its permittivity, conductivity, and thickness. This coupling creates a landscape where numerous suboptimal solutions cluster around the global optimum \cite{ref18}, significantly reducing the algorithm's efficiency in the final stages of convergence. To address this challenge and balance global exploration with local exploitation, we improved NSGA-II using the Lamarckian evolutionary algorithm and proposed an adaptive hybrid optimization algorithm based on population state awareness \cite{ref27}. By introducing a stagnation detection mechanism to dynamically control the intervention of local search, this algorithm integrates the global search capability of NSGA-II with gradient-based optimization methods. Experimental results demonstrate that, under identical initial conditions, this algorithm substantially enhances optimization performance for electromagnetic parameter extraction. Its global search capability significantly outperforms both gradient descent and traditional GA, while the incorporation of local gradient optimization ensures a faster convergence speed compared to the NSGA-II algorithm.

In the specific implementation, the algorithm employs real-number encoding, constraining the variables within predefined lower and upper bounds. The evolutionary process generates offspring via randomized selection, crossover, and mutation operators, where the mutation step size is dynamically scaled according to the search space range to effectively explore the solution space and maintain population diversity. Furthermore, its elitist strategy ensures that superior solutions are preserved throughout the process. This is achieved by merging the parent and newly generated offspring populations into a comprehensive pool, which subsequently undergoes non-dominated sorting and crowding distance evaluation. By deterministically truncating this combined population to its original size, the algorithm systematically retains the most optimal and diverse individuals for the next generation, thereby steadily guiding the search toward a well-distributed Pareto front.

To avoid inefficient gradient calculations during the early stages of algorithmic evolution, an indicator $D_t$ based on the displacement of the Pareto front's centroid is defined. When $D_t$ remains below a preset threshold $D$ for $k$ consecutive generations, the algorithm is considered to have stagnated in the neighborhood of the optimal solution. At generation $t$, the centroid $C_t$ of the Pareto front is defined as the arithmetic mean of the objective vectors of all individuals in the set, given~by

\begin{equation}
\label{centroid}
C_t\ =\ \frac{1}{\left|\mathcal{P}_t\right|}\ \sum_{x\ \in\ \mathcal{P}_t}\mathbf{F}\left(x\right)\\,
\end{equation}

\begin{equation}
\label{centroid2}
D_t=|C_t-C_{t-1}|,
\end{equation}
where $\mathcal{P}_t$ denotes the Pareto non-dominated solution set at generation $t$, $|\mathcal{P}_t |$ represents the number of solutions in the set, and $\mathbf{F}\left(x\right)$ denotes the function vector of solution in the objective space.

At this point, a local search based on gradient descent is performed on the elite solutions on the first Pareto front of each generation. Since gradient descent is only applicable to single-objective optimization, random weight scalarization is employed to temporarily transform the multi-objective optimization into single-objective subproblems

\begin{equation}
\label{centroid3}
\min_{x} \mathcal{L}(x) = \sum_{i=1}^{m} w_i \cdot f_i(x), \quad \text{s.t. } \sum_{i=1}^{m} w_i = 1.
\end{equation}

The weights $w_i$ are randomly generated, allowing different individuals to perform fine-grained searches toward various regions of the Pareto front, thereby enhancing local optimization accuracy while maintaining overall population diversity. Upon completion of the local optimization, the original individual's genes are directly replaced by those of the new individual, serving as the basis for the next generation's evolution. When the individuals optimized via local gradients show no significant improvement over the original individuals for several consecutive generations, the algorithm is considered to have converged and optimization is terminated. The algorithmic process is illustrated in Fig.~\ref{fig:flowchart}.

Compared to calculations using only reflection or transmission coefficients independently, the simultaneous use of multiple S-parameters provides a more reliable estimation of permittivity \cite{ref12}. To enhance the accuracy and physical consistency of parameter extraction, the optimization algorithm is constructed by selecting transmission measurements from two different thicknesses of the same material and reflection measurements from one of those thicknesses. The $\epsilon$, $c$, $d$ and $h$ of the two samples are set as optimization variables. This approach imposes multi-dimensional joint constraints on the transmission coefficients, reflection coefficients, and material thicknesses. The optimization is driven by an objective function defined as the root mean square error (RMSE) between the theoretical predictions and actual measurements. Specifically, the error components for the different parameters over the 20 GHz--40 GHz frequency range are given by

\begin{equation}
\label{rmse1}
\text{RMSE}_{\text{T1}} = \sqrt{\frac{1}{N} \sum_{f=\text{20 GHz}}^{\text{40 GHz}} || T_{\text{theo1}}(f) - T_{\text{meas1}}(f) ||^2},
\end{equation}

\begin{equation}
\label{rmse2}
\text{RMSE}_{\text{T2}} = \sqrt{\frac{1}{N} \sum_{f=\text{20 GHz}}^{\text{40 GHz}} || T_{\text{theo2}}(f) - T_{\text{meas2}}(f) ||^2},
\end{equation}

\begin{equation}
\label{rmse3}
\text{RMSE}_{\text{R2}} = \sqrt{\frac{1}{N} \sum_{f=\text{20 GHz}}^{\text{40 GHz}} || R_{\text{theo2}}(f) - R_{\text{meas2}}(f) ||^2},
\end{equation}
where $N$ represents the number of frequency sampling points within the specified band.

Finally, the solution with the minimum total error that is also located in a sparse region of the Pareto front is selected as the final extraction result, thereby determining the complex permittivity and thickness of the MUT.

\begin{figure*}[!t]
\captionsetup[subfloat]{labelfont={rm}, textfont={rm}}
\centering
\subfloat[]{\includegraphics[width=0.32\textwidth]{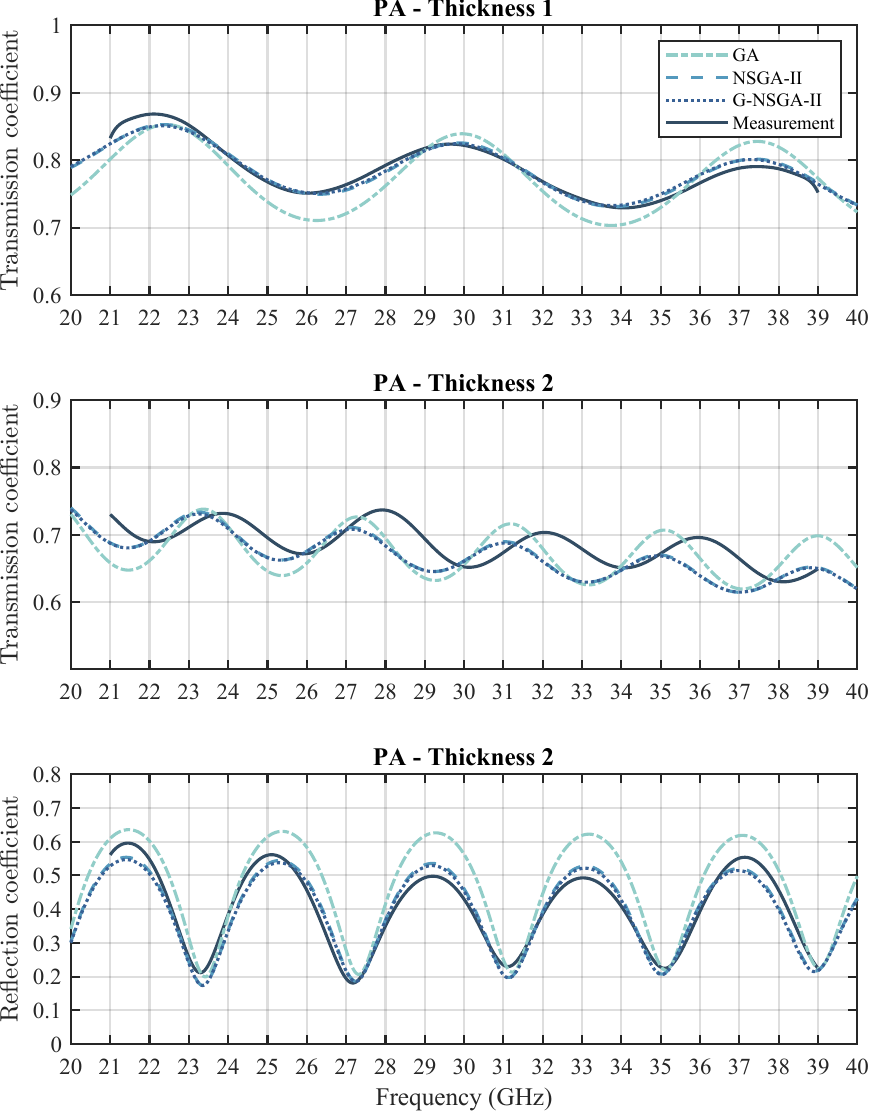}\label{fig_case1}}\hfil
\subfloat[]{\includegraphics[width=0.32\textwidth]{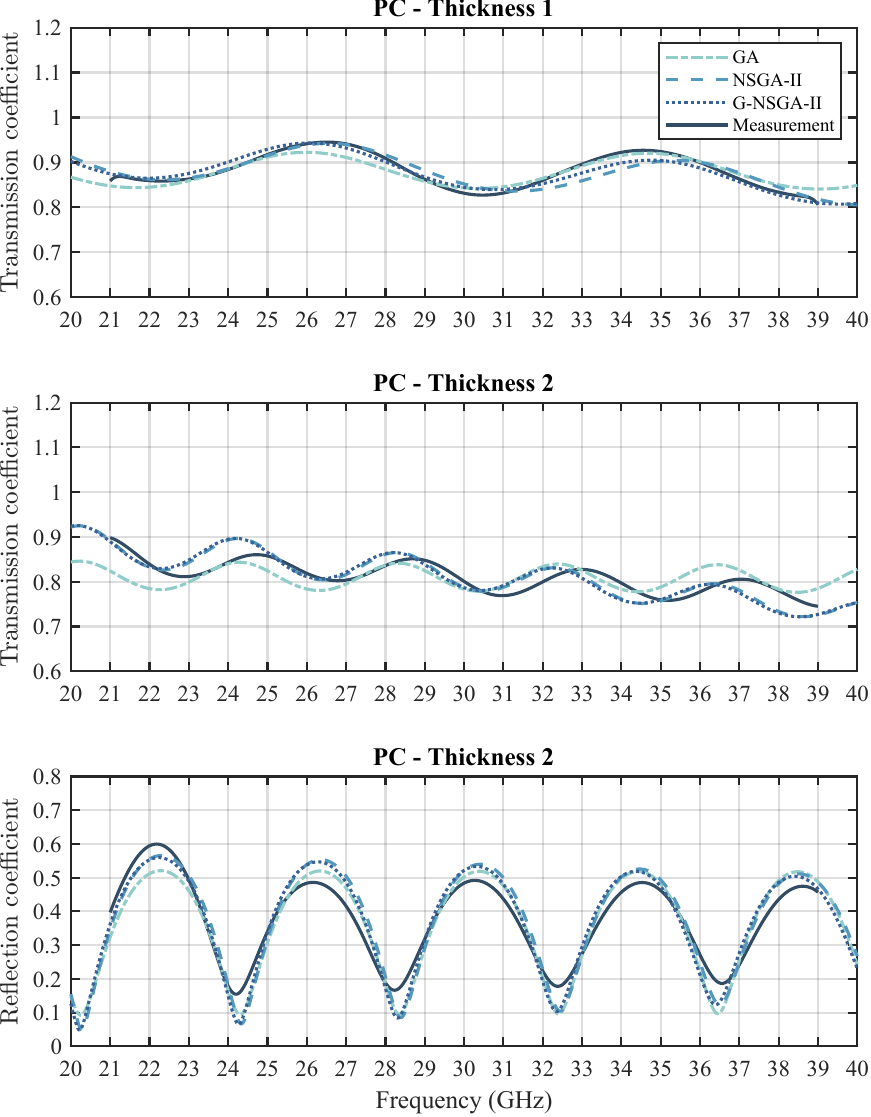}\label{fig_case2}}\hfil
\subfloat[]{\includegraphics[width=0.32\textwidth]{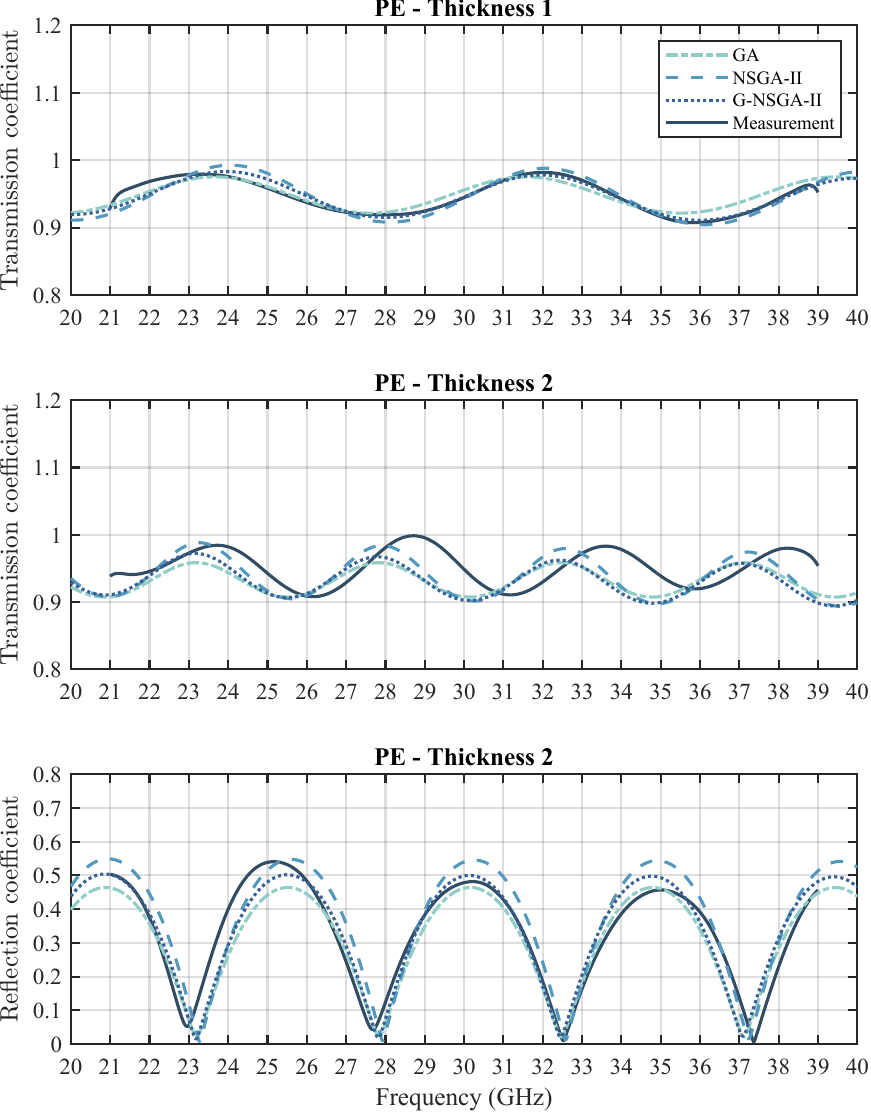}\label{fig_case3}}

\vspace{-2ex} 

\subfloat[]{\includegraphics[width=0.32\textwidth]{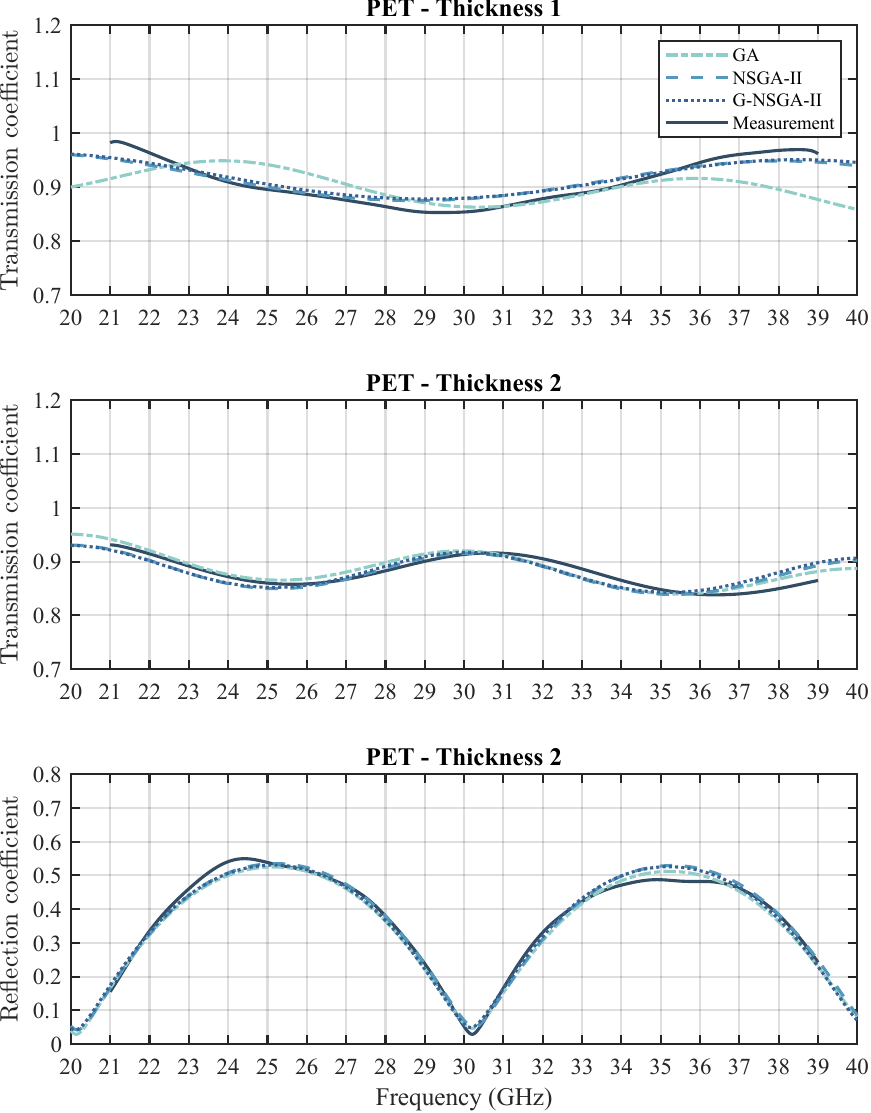}\label{fig_case4}}\hfil
\subfloat[]{\includegraphics[width=0.32\textwidth]{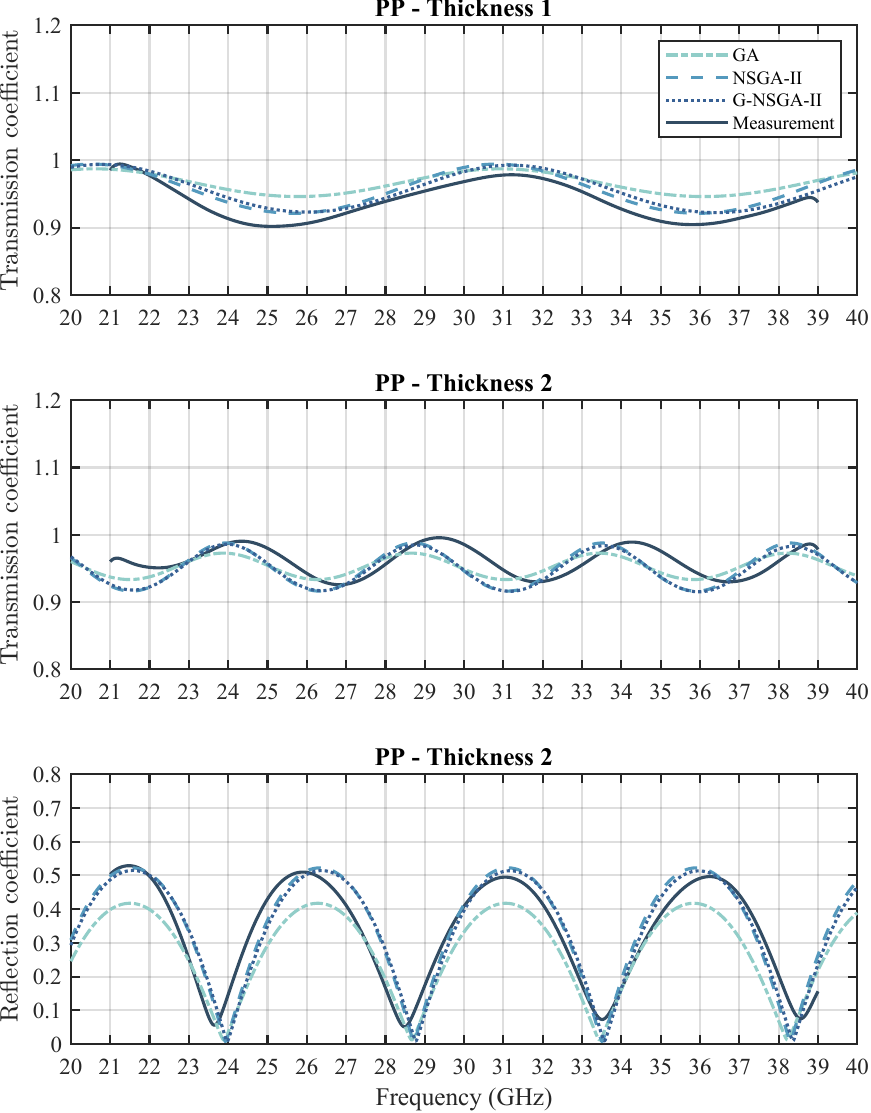}\label{fig_case5}}\hfil
\subfloat[]{\includegraphics[width=0.32\textwidth]{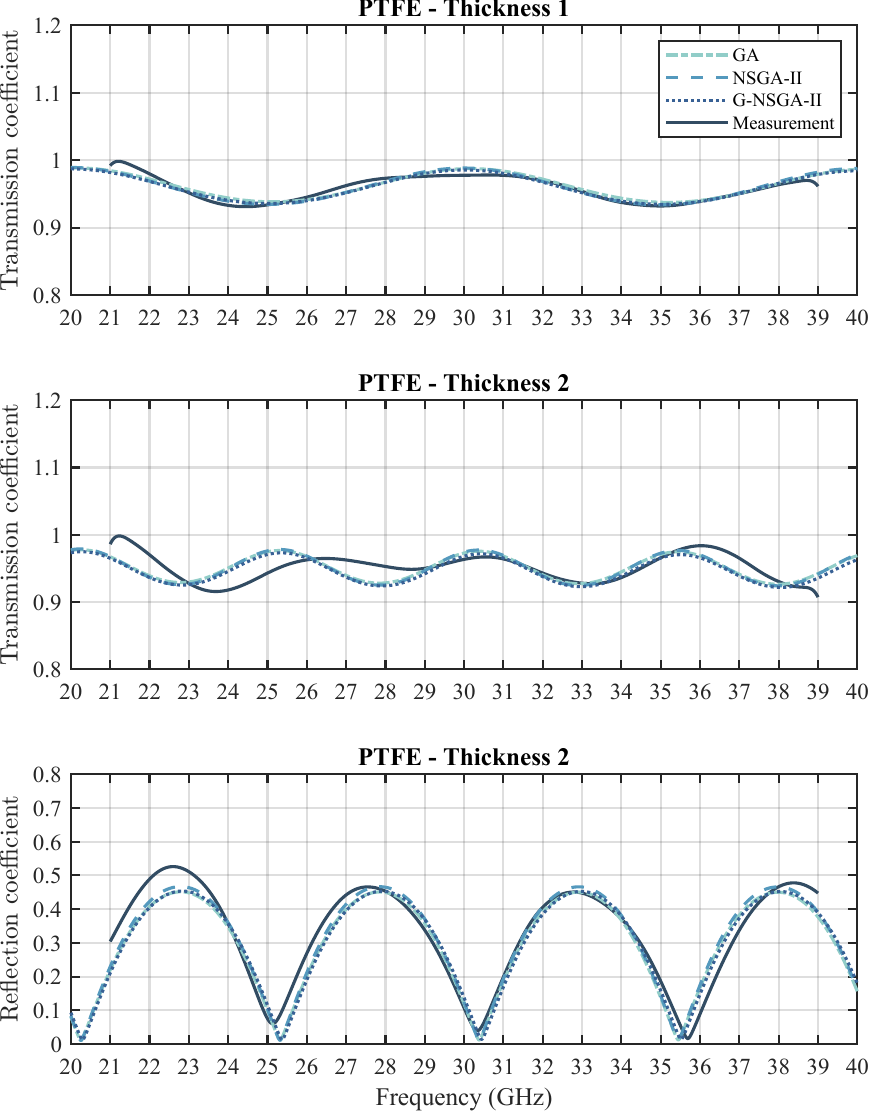}\label{fig_case6}}

\vspace{-1.5ex}

\caption{Comparison between measured and theoretical transmission and reflection coefficients for two different slab thicknesses: (a) PA, (b) PC, (c) PE, (d) PET, (e) PP, and (f) PTFE.}
\label{fig_six_images_composite}
\end{figure*}

\begin{table}[h]
\caption{Comparison of Permittivity and Measurement Methods}
\label{tab:material_result_comparison}
\centering
\setlength{\tabcolsep}{4pt} 
\begin{tabular}{ccccc}
\toprule 
\textbf{Material} & \makecell{\textbf{Relative} \\ \textbf{permittivity}} & \makecell{\textbf{Literature} \\ \textbf{values}} & \textbf{Sources} & \textbf{Method} \\
\midrule 
PA   & 2.9281 & 2.991--2.993 & [28] & Free-space \\
PC   & 2.9082 & 2.82--2.89   & [15] & Free-space \\
PE   & 2.4752 & 2.25--2.40   & \makecell{[14], [28]} & \makecell{Resonant, Free-space} \\
PET  & 2.9356 & 2.87--2.87   & [29] & Coaxial Line \\
PP   & 2.3399 & 2.25--2.30   & \makecell{[14], [28]} & \makecell{Resonant, Free-space} \\
PTFE & 2.0696 & 1.88--2.07   & \makecell{[14], [28]} & \makecell{Resonant, Free-space} \\
\bottomrule %
\end{tabular}
\end{table}

\begin{figure}[!t]
    \captionsetup[subfloat]{labelfont={rm}, textfont={rm}, margin=0pt}
    \centering
    \newcommand{\subfigwidth}{0.31\linewidth} 

    \subfloat[PA-1]{\includegraphics[width=\subfigwidth]{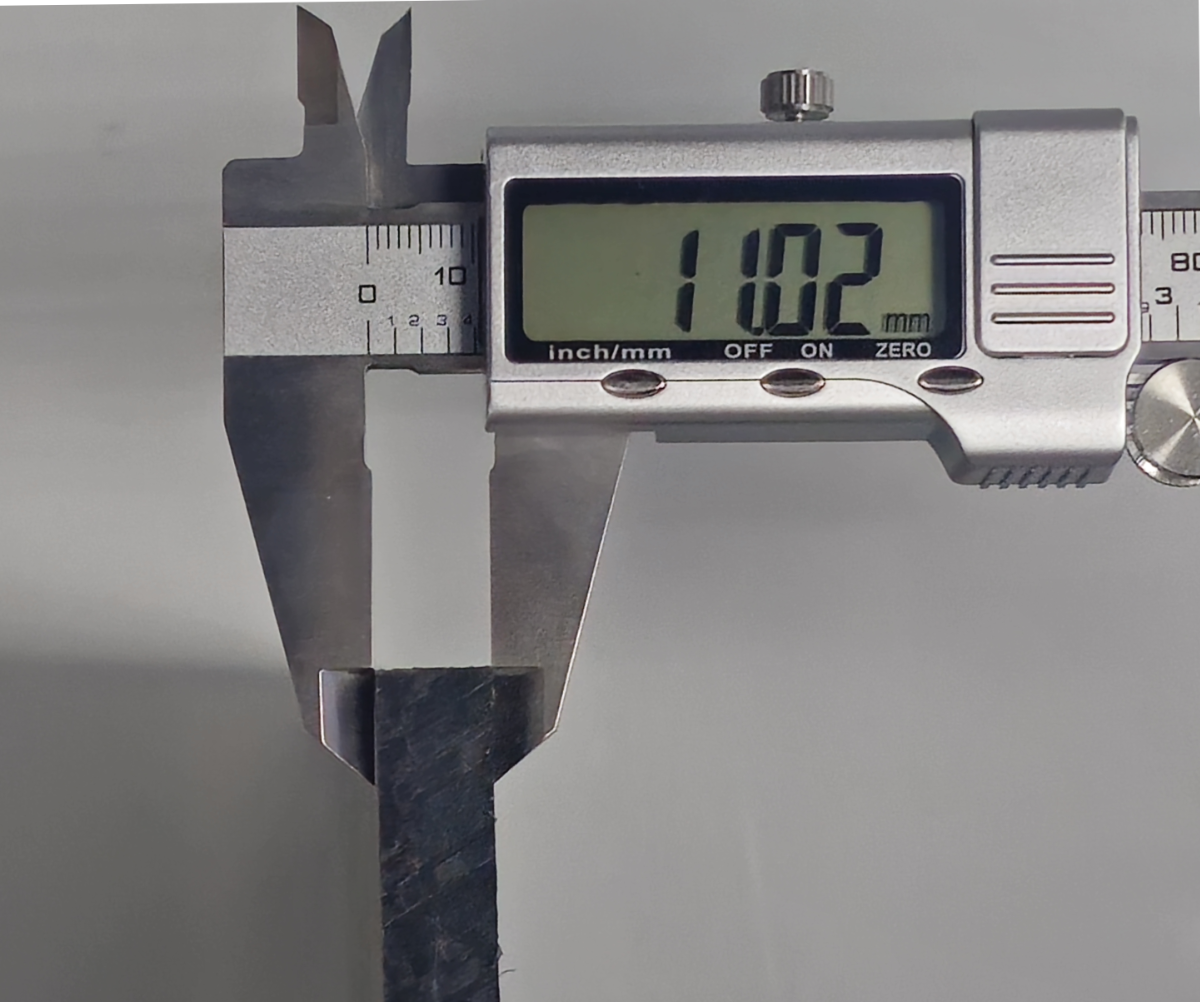}}\hfil
    \subfloat[PA-2]{\includegraphics[width=\subfigwidth]{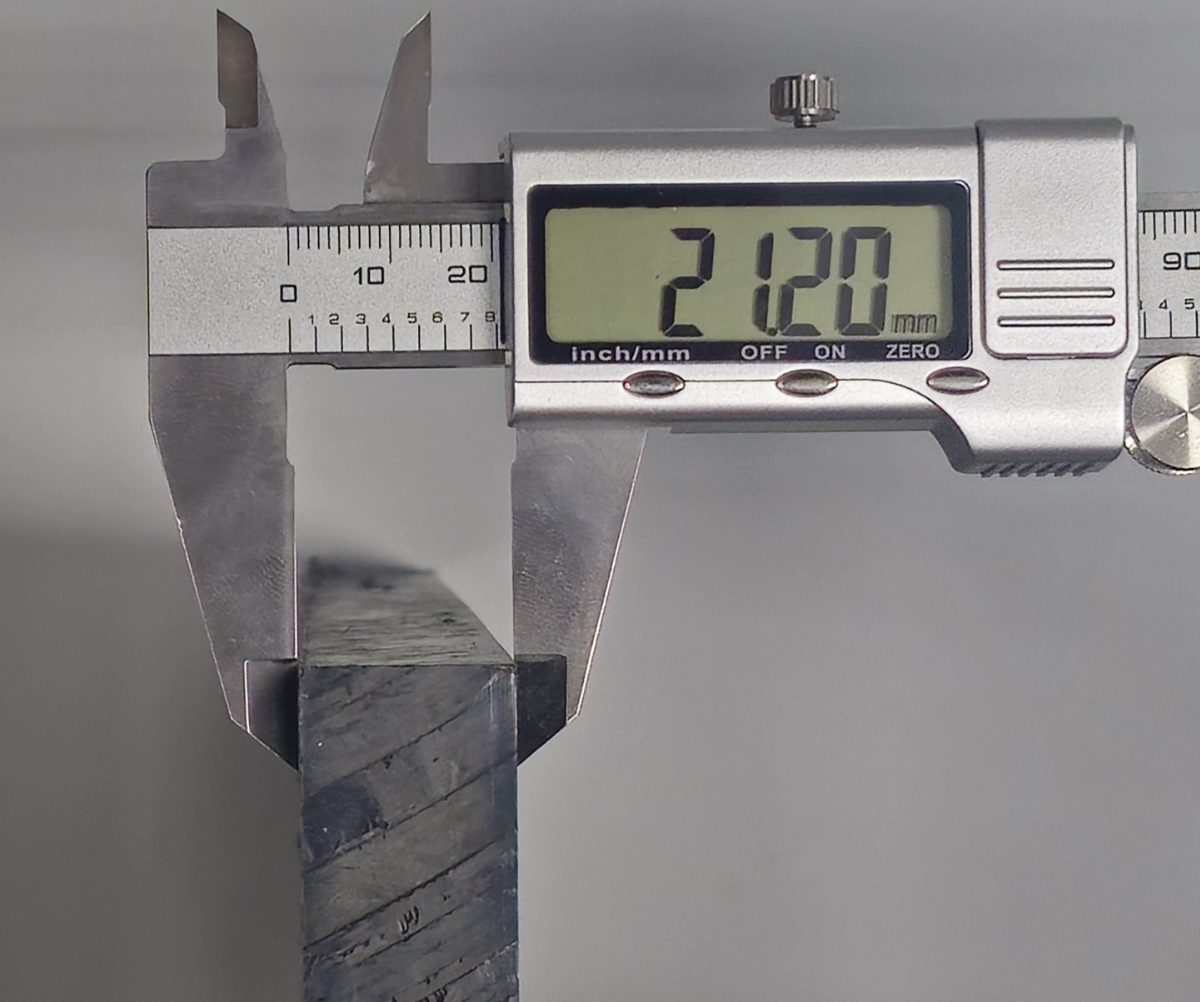}}\hfil
    \subfloat[PC-1]{\includegraphics[width=\subfigwidth]{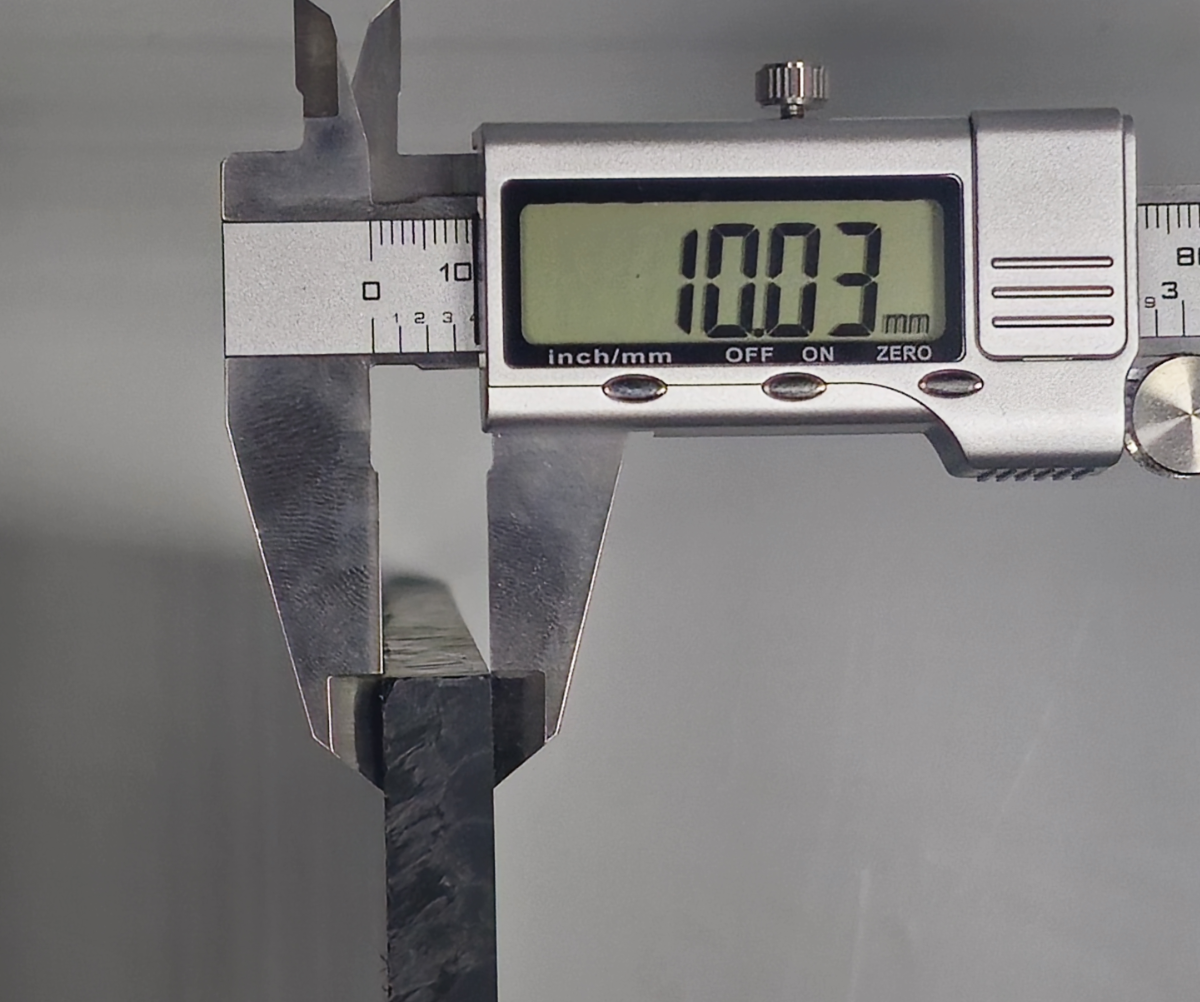}}
    \vspace{-1ex}

    \subfloat[PC-2]{\includegraphics[width=\subfigwidth]{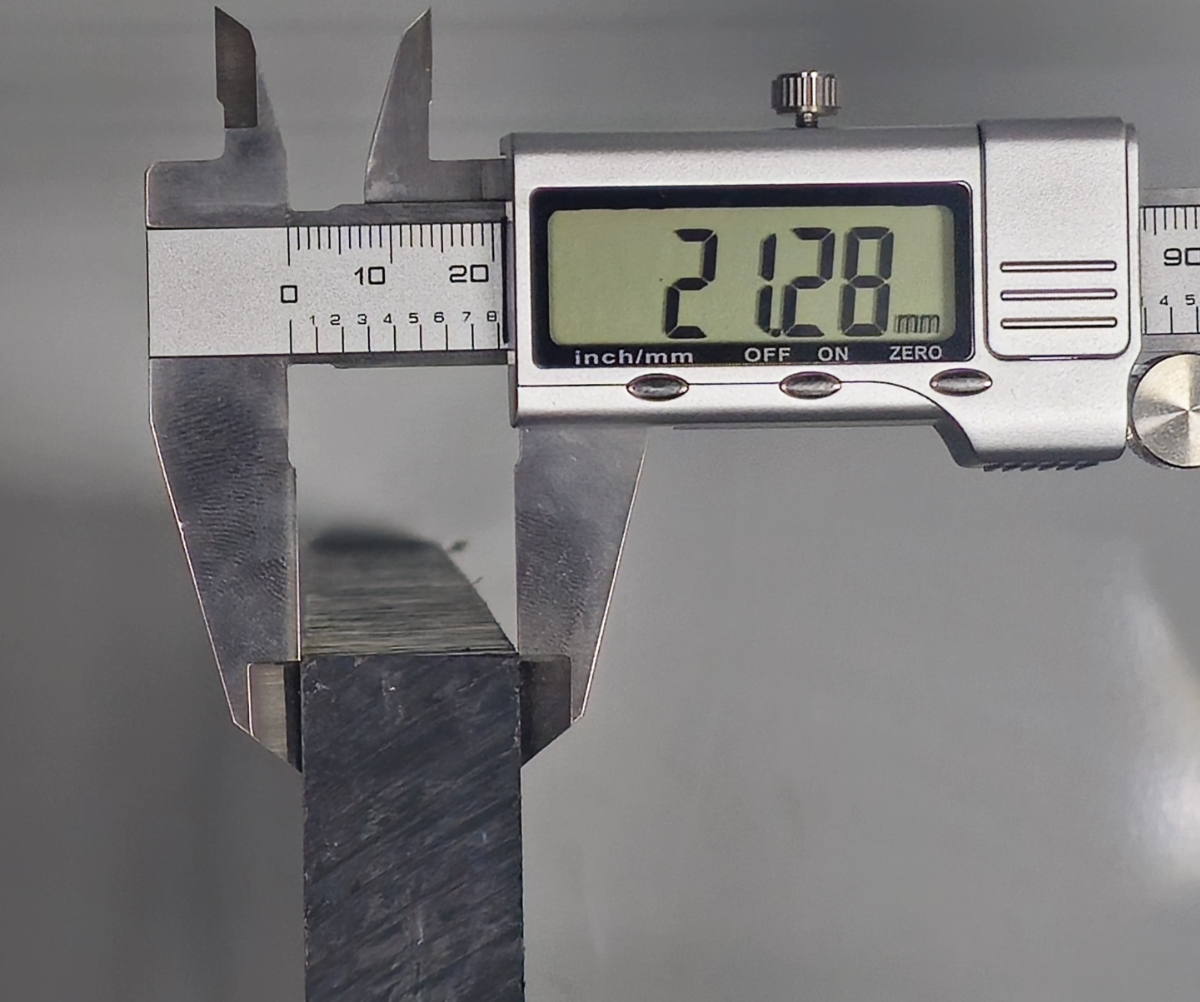}}\hfil
    \subfloat[PE-1]{\includegraphics[width=\subfigwidth]{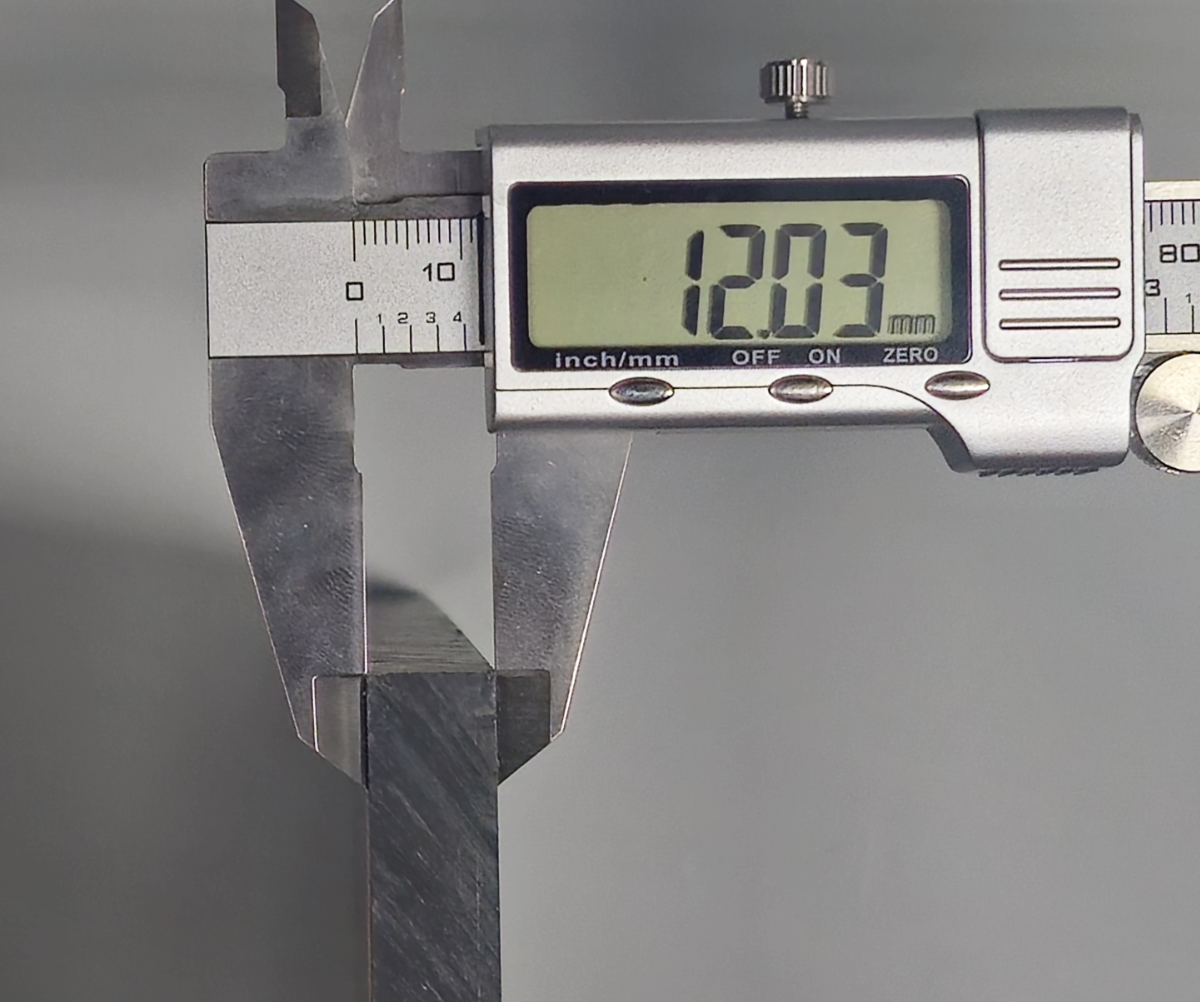}}\hfil
    \subfloat[PE-2]{\includegraphics[width=\subfigwidth]{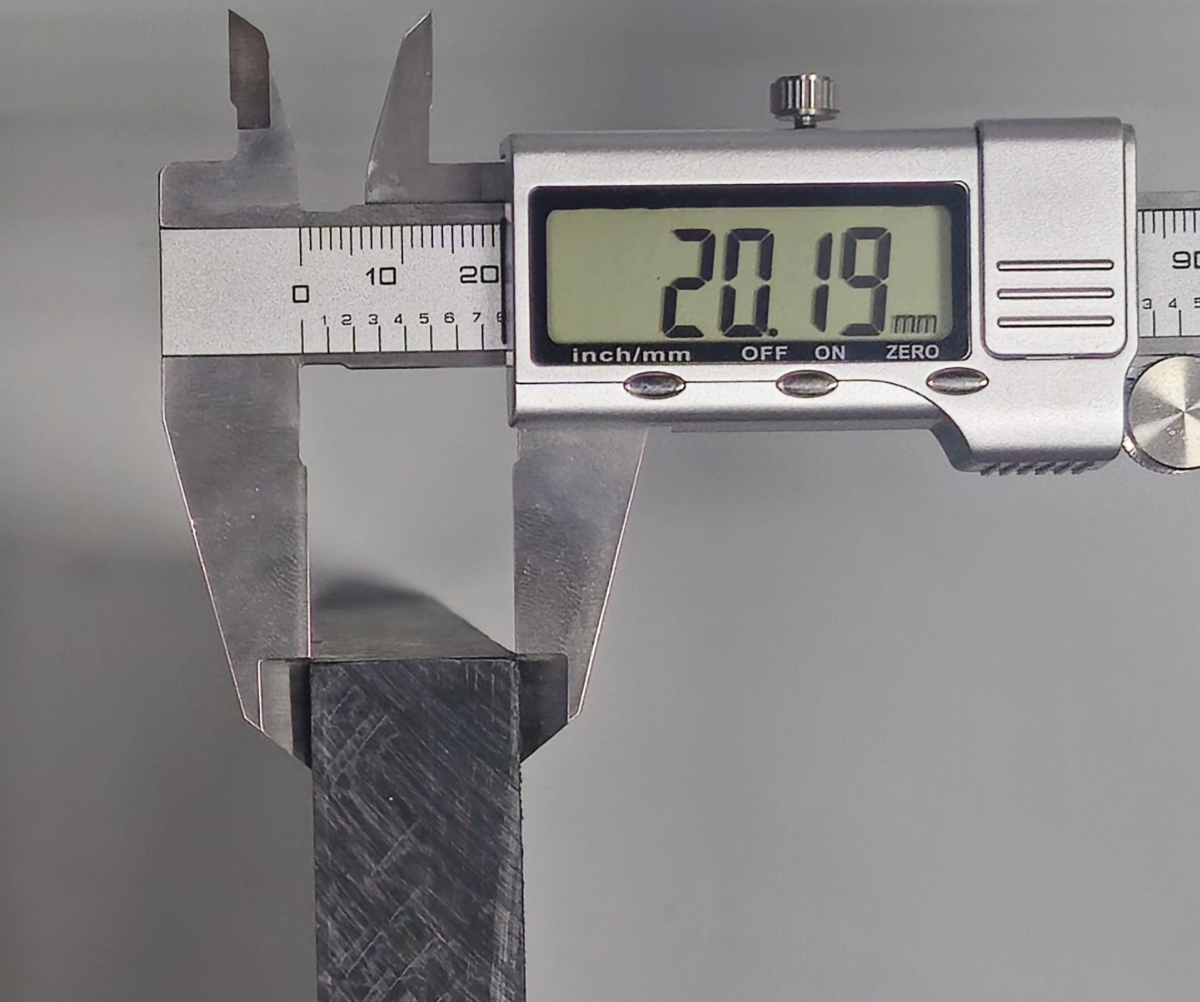}}
    \vspace{-1ex}

    \subfloat[PET-1]{\includegraphics[width=\subfigwidth]{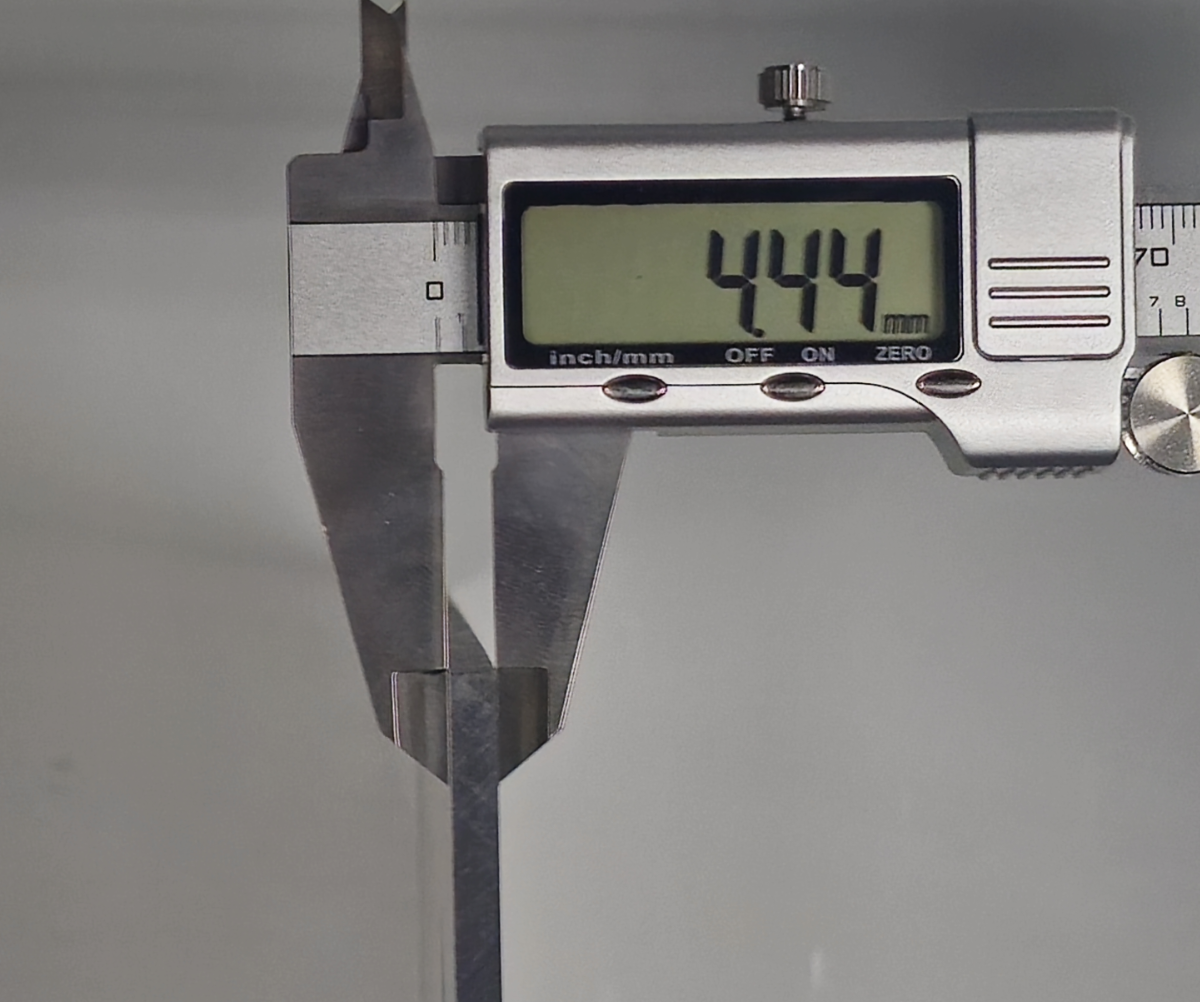}}\hfil
    \subfloat[PET-2]{\includegraphics[width=\subfigwidth]{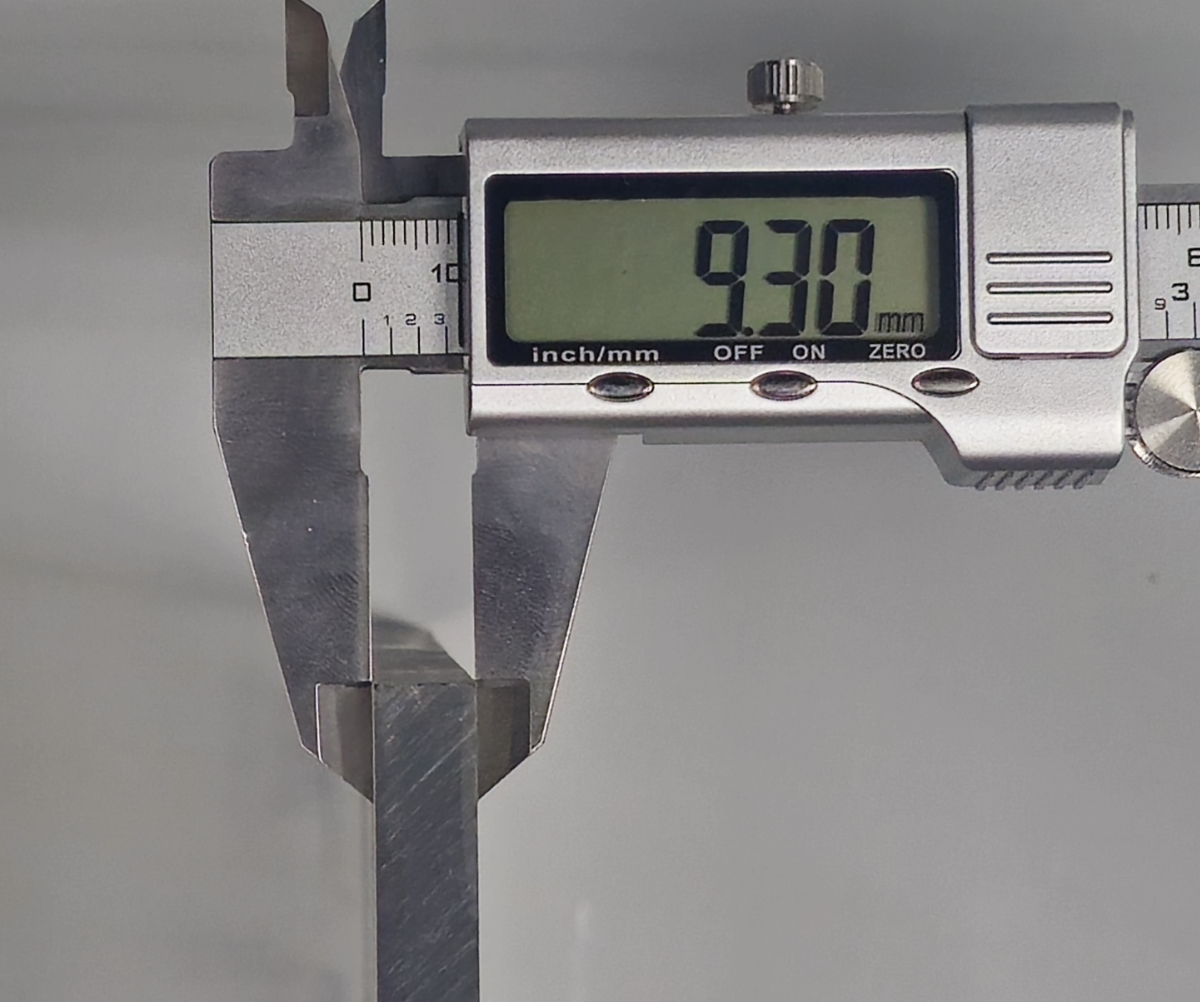}}\hfil
    \subfloat[PP-1]{\includegraphics[width=\subfigwidth]{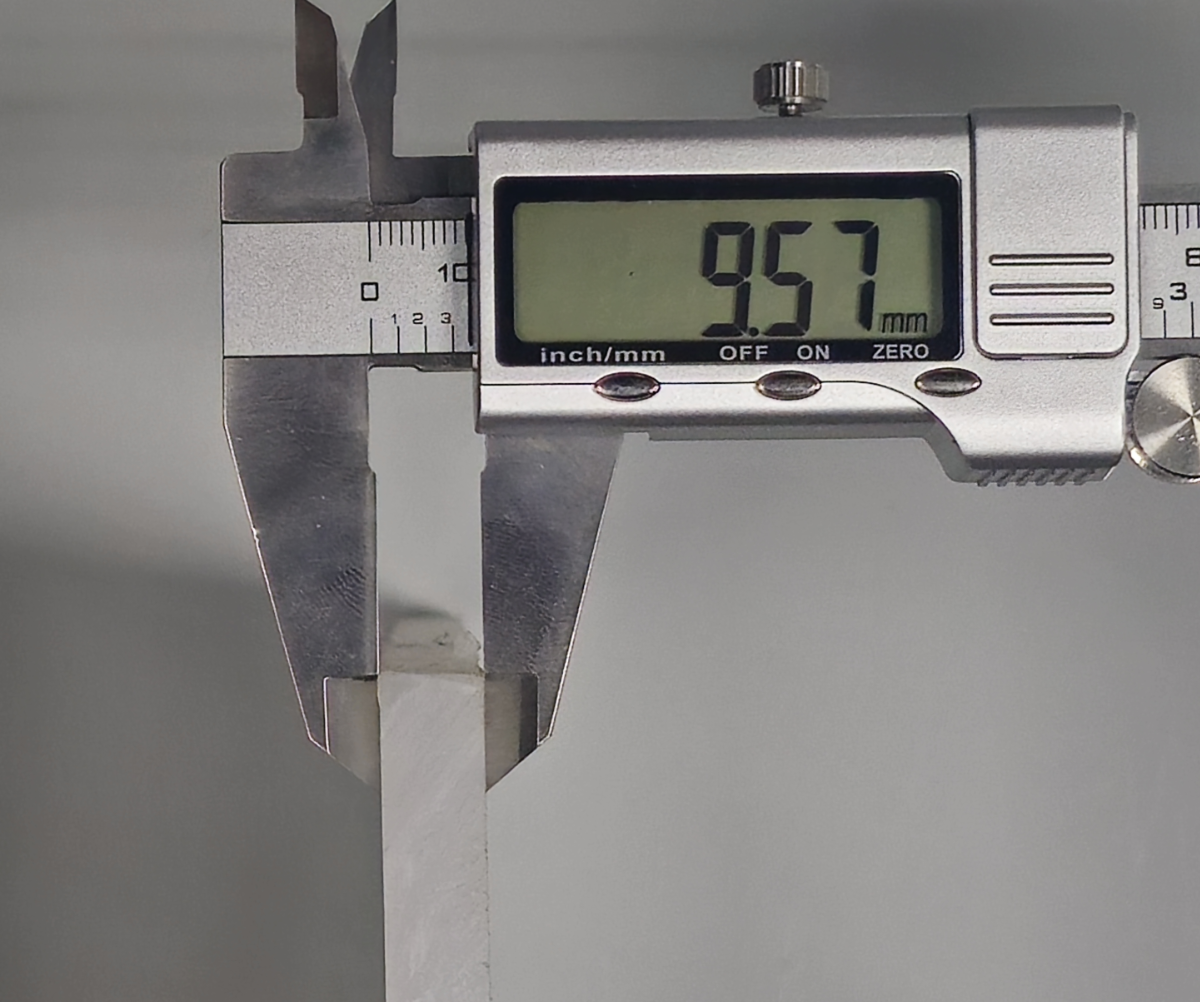}}
    \vspace{-1ex}

    \subfloat[PP-2]{\includegraphics[width=\subfigwidth]{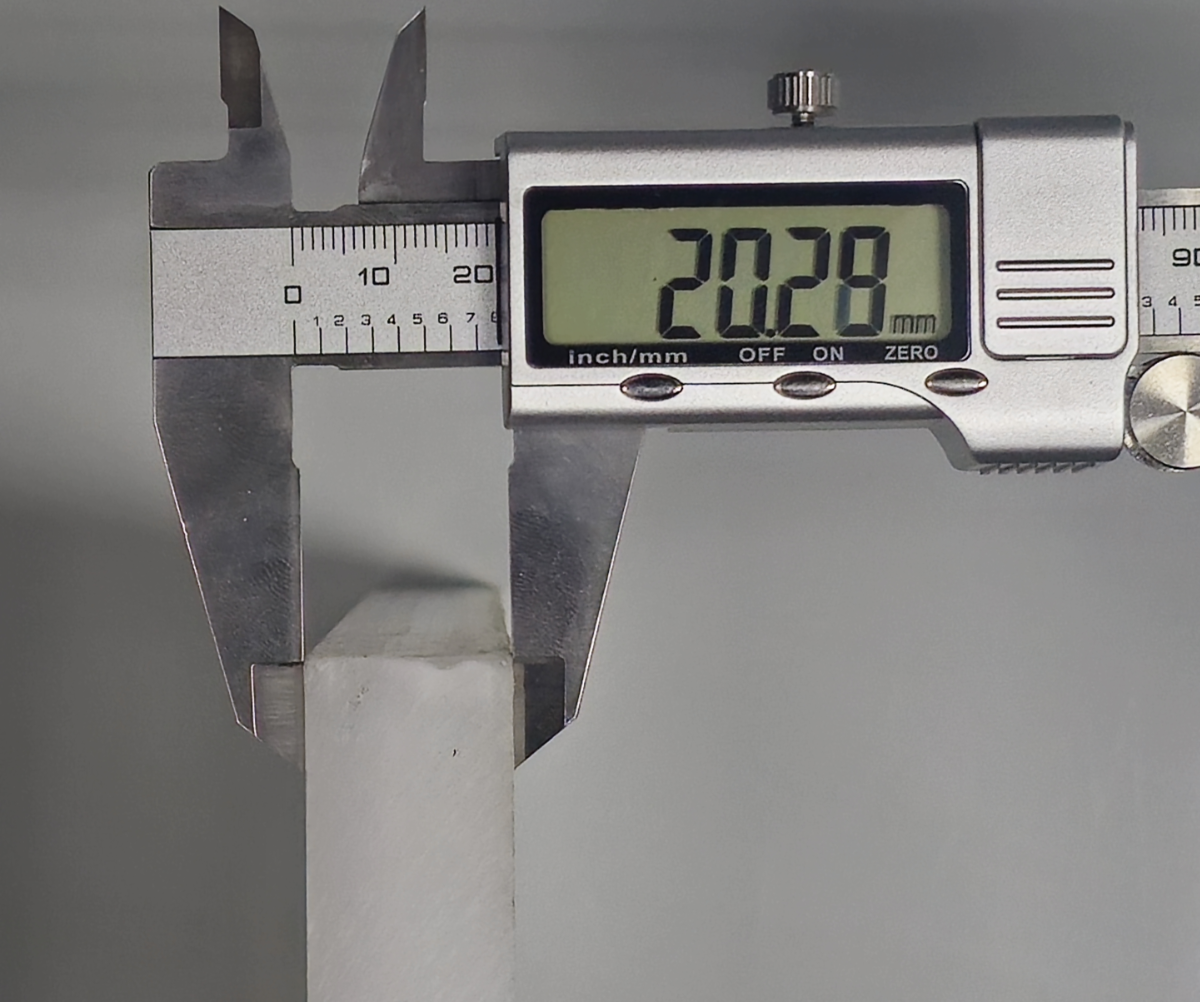}}\hfil
    \subfloat[PTFE-1]{\includegraphics[width=\subfigwidth]{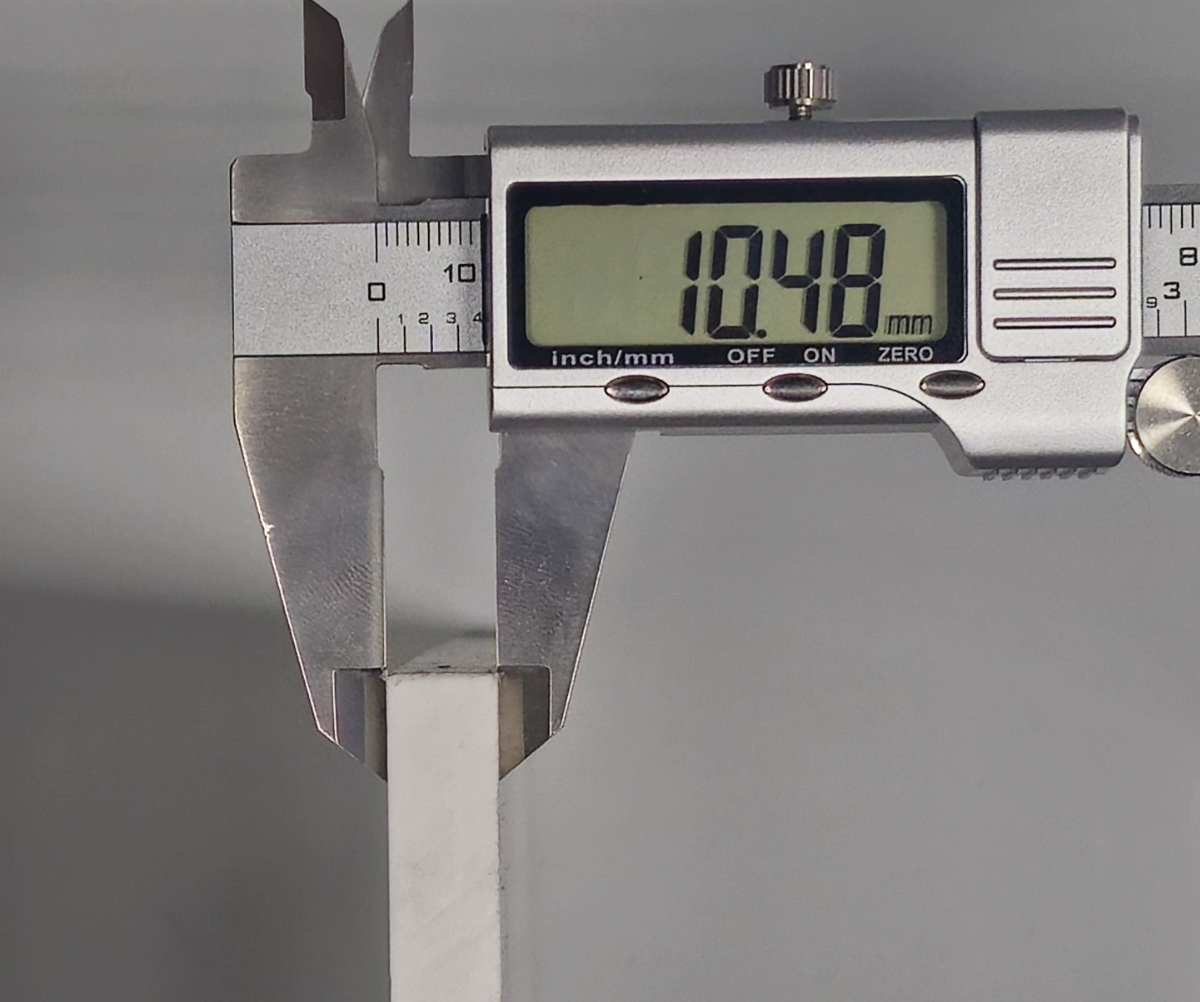}}\hfil
    \subfloat[PTFE-2]{\includegraphics[width=\subfigwidth]{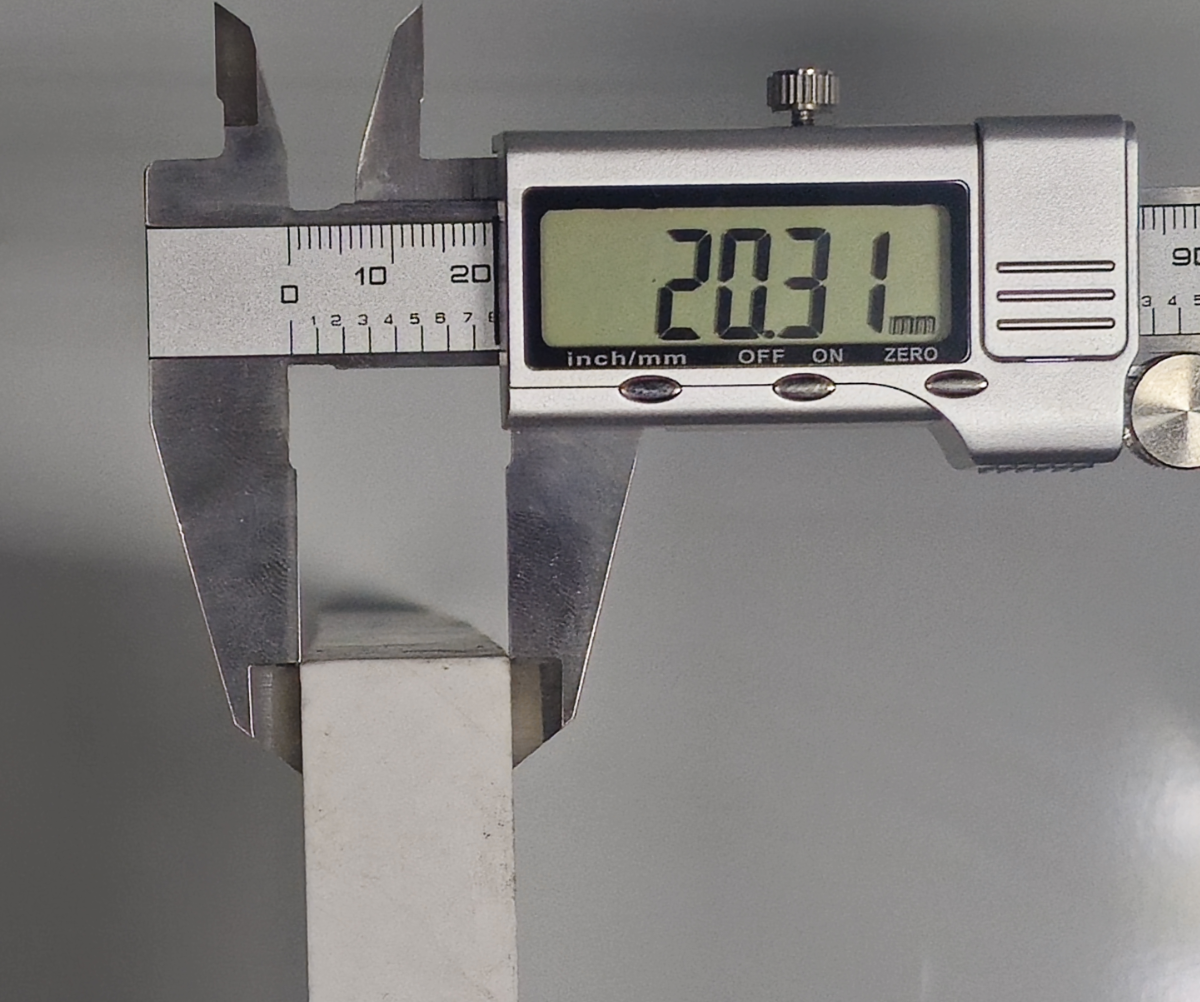}}

    \caption{Manual measurement of slab thicknesses using a vernier caliper for the six polymer slabs (PA, PC, PE, PET, PP, and PTFE) at two distinct thicknesses.}
    \label{fig:material_comparison}
    \vspace{-2ex}
\end{figure}

\section{Result and Discussion}

The relative permittivity of the polymer materials, derived from our measurements and the proposed extraction calculations, is presented in Table~\ref{tab:material_result_comparison}. To quantitatively evaluate the measurement accuracy this paper adopts the arithmetic mean of reference ranges from the literature as a benchmark to calculate the percentage error for each material.

The calculated relative permittivity slightly deviates from certain reference studies. These discrepancies are primarily attributed to manufacturing variations in the actual measured samples, such as differences in raw material purity, rather than methodological flaws. Despite these physical variations, the measurement errors for the majority of materials remain at a remarkably low level. Error rates for PA, PET, and PC were notably low (2.29\%, 2.14\% and 1.86\%), whereas those for PE, PP, and PTFE were 6.46\%, 2.85\%, and 4.79\%, respectively. These results corroborate that the proposed method effectively mitigates the challenges of high-precision electromagnetic parameter extraction across a wide frequency spectrum, fulfilling the requirement for accurate indoor channel simulation.

To address the solution non-uniqueness often encountered in traditional free-space methods, this study incorporated multi-dimensional constraints by optimizing across multiple sample thicknesses. Fig.~\ref{fig_six_images_composite} presents the measured transmission and reflection coefficients for two different plate thicknesses of each material, alongside the values calculated by the algorithm proposed in this paper. The theoretical values derived from the model demonstrate consistent agreement with the measured data across the entire 20-40 GHz frequency band. This rigorous alignment validates our strategy of jointly optimizing coefficients across multiple thicknesses. This approach ensures both mathematical uniqueness and physical reliability in extracting broadband electromagnetic parameters, thereby overcoming the limitations of single-thickness or reflection-only methods.

\begin{figure}[!t]
    \centering
    \captionsetup[subfloat]{labelfont={rm}, textfont={rm}}
    \subfloat[Transmission 1]{\includegraphics[width=0.15\textwidth]{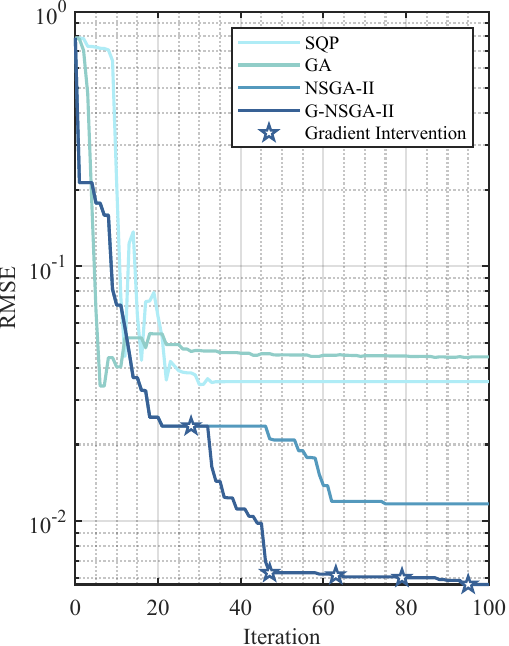}\label{fig:obj1}}
    \hfill
    \subfloat[Transmission 2]{\includegraphics[width=0.15\textwidth]{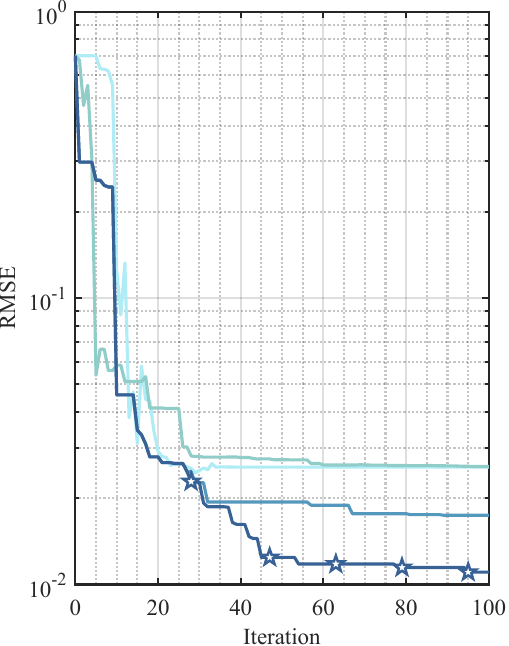}\label{fig:obj2}}
    \hfill
    \subfloat[Reflection 2]{\includegraphics[width=0.15\textwidth]{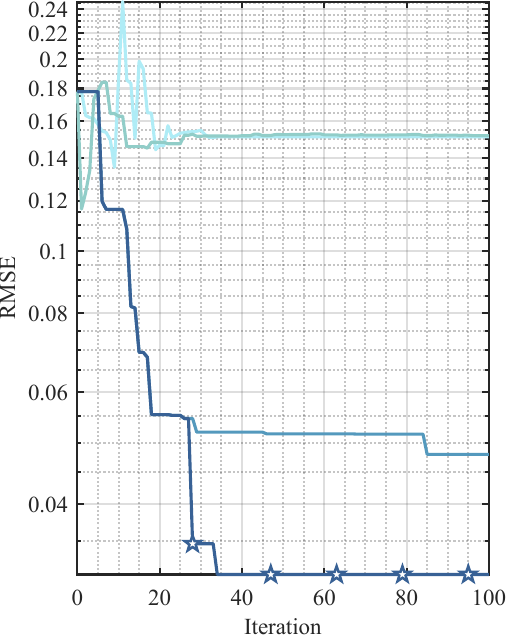}\label{fig:obj3}}
    
    \caption{Convergence characteristics of SQP, GA, NSGA-II, and the proposed G-NSGA-II starting from identical initial conditions.}
    \label{Convergenccompare}
\end{figure}

A key contribution of this work is the adaptive hybrid strategy designed to overcome the inherent ill-posedness of inverse problems. Fig.~\ref{Convergenccompare} compares the practical performance of sequential quadratic programming (SQP), GA, NSGA-II, and our proposed G-NSGA-II for the electromagnetic parameter extraction problem. The comparative analysis reveals distinct performance characteristics: 
\begin{itemize} 
    \item SQP: Under identical initial conditions, population sizes, and genetic configurations, the SQP algorithm quickly became trapped in a local optimum and ceased its search, clearly illustrating the susceptibility of traditional gradient methods to the non-convex nature of the objective function.
    \item GA: Due to a lack of multi-dimensional constraints and limited local search capabilities, the GA algorithm frequently fell into local optima, demonstrating weaker robustness compared to the multi-objective approaches. 
    \item NSGA-II: While the NSGA-II algorithm exhibits superior multi-objective global search capabilities and approaches the global optimum effectively, its weak local search capability prevents it from converging to the exact global optimal solution. NSGA-II converges rapidly in the first 30 iterations but decelerates significantly as it approaches the solution space, suffering from the slow precision improvement issue previously mentioned in the introduction.
    \item G-NSGA-II (Proposed): The G-NSGA-II algorithm successfully balances global exploration with local refinement. The star symbol in Fig.~\ref{Convergenccompare} marks the intervention point of the local gradient search. Once the neighborhood of the global optimum is reached, the algorithm introduces gradient search. With just two rounds of intervention, G-NSGA-II achieves convergence by the 50th generation. This demonstrates that the proposed hybrid strategy not only enhances convergence rates but also guarantees algorithmic robustness, effectively solving the optimization challenges that limit standard evolutionary frameworks. 
\end{itemize}

\begin{table}[t]
\caption{Comparison of Actual and Calculated Thicknesses}
\label{tab:material_data}
\centering
\begin{tabular}{ccccc}
\toprule 
\textbf{Material} & \makecell{\textbf{Actual} \\ \textbf{$h_1$ (mm)}} & \makecell{\textbf{Calculated} \\ \textbf{$h_1$ (mm)}} & \makecell{\textbf{Actual} \\ \textbf{$h_2$ (mm)}} & \makecell{\textbf{Calculated} \\ \textbf{$h_2$ (mm)}} \\
\midrule 
PA   & 11.02 & 11.700 & 21.20 & 21.884 \\
PC   & 10.03 & 10.048 & 21.28 & 21.291 \\
PE   & 12.03 & 12.042 & 20.19 & 20.023 \\
PET  & 4.44  & 4.485  & 9.30  & 9.362  \\
PP   & 9.57  & 9.638  & 20.28 & 20.298 \\
PTFE & 10.48 & 10.486 & 20.31 & 20.108 \\
\bottomrule 
\end{tabular}
\end{table}

Finally, to validate the non-destructive and calibration-free advantages of the proposed method, we evaluated the algorithm's ability to invert sample thickness simultaneously with electromagnetic parameters. Fig.~\ref{fig:material_comparison} illustrates the manual measurement process for the six polymer slabs (PA, PC, PE, PET, PP, and PTFE) using a vernier caliper. Subsequently, Table~\ref{tab:material_data} presents a comparison between these actual measured thicknesses and the values calculated by the optimization algorithm. The results indicate high precision, with an average error of 1.3719\% for materials with thickness $h_1$ and 0.9759\% for thickness $h_2$, resulting in an overall average error of 1.1739\%. These minimal discrepancies across a diverse range of dielectric profiles underscore the algorithm's capacity to reliably decouple physical dimensions from constitutive parameters. The G-NSGA-II algorithm not only extracts the electromagnetic parameters of MUT with high precision and robustness but also accurately determines the MUT thickness. This capability significantly simplifies the operational workflow of the free-space method. By eliminating the need for precise pre-measurement machining or rigorous physical thickness measurements, the algorithm allows thickness to be treated as an optimization parameter. This effectively addresses the rigorous sample preparation limitation associated with waveguide and resonant cavity techniques, making the method highly suitable for practical applications in building material characterization.

\section{Conclusions}
Through experimental validation on six typical polymers, this paper demonstrates the outstanding performance of the proposed G-NSGA-II in extracting material electromagnetic parameters, while simultaneously determining the MUT thickness with high precision. By combining transmission coefficients, reflection coefficients, and thicknesses joint optimization for materials of varying thickness, the method effectively overcomes the common issue of becoming trapped in local optima prevalent in traditional extraction algorithms. The complex permittivity extracted in this study is in high agreement with literature reference values. Furthermore, comparative experiments over 100 generations demonstrate the superior efficiency of the proposed G-NSGA-II. While traditional gradient and heuristic methods failed to fully converge or became trapped in local optima within the full cycle, G-NSGA-II achieved stable convergence in just 50 generations. This significant improvement in speed, without compromising precision, offers a highly reliable solution for the wideband parameter characterization of materials.

\vfill


\begin{thebibliography}{1}
\bibliographystyle{IEEEtran}

\bibitem{populor}
T. S. Rappaport, Y. Xing, G. R. MacCartney, A. F. Molisch, E. Mellios and J. Zhang, ``Overview of Millimeter Wave Communications for Fifth-Generation (5G) Wireless Networks—With a Focus on Propagation Models," in \textit{IEEE Trans. Antennas Propag.}, vol. 65, no. 12, pp. 6213-6230, Dec. 2017.

\bibitem{bwp1}
Y. Zhou, Y. Shao, J. Zhang and J. Zhang, ``Wireless Performance Evaluation of Building Materials Integrated With Antenna Arrays," in \textit{IEEE Commun. Lett.}, vol. 26, no. 4, pp. 942-946, Apr. 2022.

\bibitem{Persad_Infill_2024}
J. Persad and S. Rocke, ``Impact of 3D Printing Infill Patterns on the Effective Permittivity of 3D Printed Substrates," in \textit{IEEE J. Microw.}, vol. 4, no. 2, pp. 277-292, Apr. 2024.

\bibitem{bwp2}
Y. Zhang, J. Zhang, X. Chu and J. Zhang, ``Wireless Friendliness Evaluation and Optimization for Sandwich Building Materials as Reflectors," in \textit{IEEE Trans. Antennas Propag.}, vol. 72, no. 3, pp. 2697-2711, Mar. 2024.

\bibitem{Li_SiC}
T. Li, L. Li, X. Wang, J. C. M. Hwang, S. Yanagimoto and Y. Yanagimoto, ``Ordinary and Extraordinary Permittivities of 4H SiC at Different Millimeter-Wave Frequencies, Temperatures, and Humidities," in \textit{IEEE J. Microw.}, vol. 4, no. 4, pp. 666-674, Oct. 2024.

\bibitem{Sanchez_Crosstalk_2026}
J. P. Sánchez-Muñoz, S. C. Sejas-García, C. Nwachukwu and R. Torres-Torres, ``Effect of Dielectric Permittivity Non-Uniformity on Microwave Far-End Crosstalk in Coupled PCB Microstrip Transmission Lines," in \textit{IEEE J. Microw.}, vol. 6, no. 1, pp. 263-272, Jan. 2026.

\bibitem{Alhassoon_Extract}
K. A. Alhassoon, Y. Malallah and A. S. Daryoush, ``Complex Permittivity and Permeability Extraction of Ferromagnetic Materials For Magnetically Tuned Microwave Circuits," in \textit{IEEE J. Microw.}, vol. 1, no. 2, pp. 639-645, Apr. 2021.

\bibitem{ref9}
A. Karatay and F. Yaman, ``Mixture-Based Dielectric Permittivity Measurements Through Gallium-Excited Cavities," in \textit{IEEE Trans. Instrum. Meas.}, vol. 73, pp. 1-8, Jul. 2024.

\bibitem{ref10}
Q. Shi, Q.-X. Chu, M.-Z. Xiao, F.-C. Chen, X.-Q. Huang and X. He, ``Complex Permittivity Measurement Utilizing Multiple Modes of a Rectangular Cavity," in \textit{IEEE Trans. Instrum. Meas.}, vol. 72, pp. 1-8, Dec. 2023.

\bibitem{ref11}
X. Han et al., ``CSRR Metamaterial Microwave Sensor for Measuring Dielectric Constants of Solids and Liquids," in \textit{IEEE Sens. J.}, vol. 24, no. 9, pp. 14167-14176, May, 2024.

\bibitem{ref12}
G. González-López, S. Blanch, J. Romeu and L. Jofre, ``Debye Frequency-Extended Waveguide Permittivity Extraction for High Complex Permittivity Materials: Concrete Setting Process Characterization," in \textit{IEEE Trans. Instrum. Meas.}, vol. 69, no. 8, pp. 5604-5613, Aug. 2020.

\bibitem{ref13}
Q. Huang et al., ``A Novel Broadband Measurement Method for Electromagnetic Parameters of Liquids Based on Resonance and Transmission-Reflection of Striplines," in \textit{IEEE Trans. Instrum. Meas.}, vol. 74, pp. 1-16, Apr. 2025.

\bibitem{ref14}
J. Krupka, ``Measurements of the Complex Permittivity of Low Loss Polymers at Frequency Range From 5 GHz to 50 GHz," in \textit{IEEE Microw. Wirel. Compon. Lett.}, vol. 26, no. 6, pp. 464-466, Jun. 2016.

\bibitem{Chang_FreeSpace}
S. -M. Chang, C. Swank, A. Kummel and J. F. Buckwalter, ``Free Space Dielectric Techniques for Diamond Composite Characterization," in \textit{IEEE J. Microw.}, vol. 4, no. 1, pp. 147-157, Jan. 2024.

\bibitem{Brandl_Rotation}
A. Diepolder, M. Mueh, S. Brandl, P. Hinz, C. Waldschmidt and C. Damm, ``A Novel Rotation-Based Standardless Calibration and Characterization Technique for Free-Space Measurements of Dielectric Material," in \textit{IEEE J. Microw.}, vol. 4, no. 1, pp. 56-68, Jan. 2024.

\bibitem{ref15}
C. A. Grosvenor, R. T. Johnk, J. Baker-Jarvis, M. D. Janezic and B. Riddle, ``Time-Domain Free-Field Measurements of the Relative Permittivity of Building Materials," in \textit{IEEE Trans. Instrum. Meas.}, vol. 58, no. 7, pp. 2275-2282, Jul. 2009.

\bibitem{ref16}
A. Rashidian, L. Shafai, D. Klymyshyn and C. Shafai, ``A Fast and Efficient Free-Space Dielectric Measurement Technique at mm-Wave Frequencies," in \textit{IEEE Trans. Antennas Propag. Lett.}, vol. 16, pp. 2630-2633, Aug. 2017.

\bibitem{ref17}
U. C. Hasar, G. Ozturk, Y. Kaya and M. Ertugrul, ``Calibration-Free Time-Domain Free-Space Permittivity Extraction Technique," in \textit{IEEE Trans. Antennas Propag.}, vol. 70, no. 2, pp. 1565-1568, Feb. 2022.

\bibitem{ref18}
B. Xue, ``Characterizing Complex Permittivity of Thin Materials by Free-Space Transmission and Reflection Coefficients," in \textit{IEEE Antennas Wireless Propagat. Lett.}, vol. 23, no. 12, pp. 4438-4442, Dec. 2024.

\bibitem{ref19}
Y. Zhang, J. Zhang, Y. Zhou, C. Gao, Y. Gao and E. Li, ``Correction of Complex Permittivity Inversion in Free-Space Gaussian Beam Reflection Model," in \textit{IEEE Trans. Antennas Propag.}, vol. 69, no. 10, pp. 6712-6722, Oct. 2021.

\bibitem{ref20}
U. C. Hasar, H. Korkmaz, Y. Kaya and T. Iliev, ``Soil Permittivity Extraction by Time-Domain Free-Space Calibration-Free Microwave Measurements With No Thickness Information," in \textit{IEEE Trans. Geosci. Remote Sens.}, vol. 63, pp. 1-13, May 2025.

\bibitem{ref21}
J. Luo, Y. Shao, X. Liao, J. Liu and J. Zhang, ``Complex Permittivity Estimation for Cloths Based on QPSO Method Over (40 to 50) GHz," in \textit{IEEE Trans. Antennas Propag.}, vol. 69, no. 1, pp. 600-605, Jan. 2021.

\bibitem{ref22}
S. Talebi and H. H. Aly, ``Optimized Renewable Energy Integration: Advanced Modeling, Control, and Design of a Standalone Microgrid Using Hybrid FA-PSO," in \textit{IEEE Access}, vol. 13, pp. 63486-63503, Apr. 2025.

\bibitem{ref23}
S. Prithi and S. Sumathi, ``LD2FA-PSO: A novel Learning Dynamic Deterministic Finite Automata with PSO algorithm for secured energy efficient routing in Wireless Sensor Network," in \textit{Ad Hoc Netw.}, vol. 97, Art. no. 102024, Feb. 2020.


\bibitem{material_antenna}
Y. Zhou, Y. Shao, J. Zhang and J. Zhang, ``Wireless Performance Evaluation of Building Materials Integrated With Antenna Arrays," in \textit{IEEE Communi. Lett.}, vol. 26, no. 4, pp. 942-946, Apr. 2022.

\bibitem{ref24}
\textit{Effects of Building Materials and Structures on Radiowave Propagation Above About 100 MHz P Series Radiowave Propagation}, Recomm. document ITU-R P.2040-1, Jul. 2015.

\bibitem{ref25}
K. Deb, A. Pratap, S. Agarwal and T. Meyarivan, ``A fast and elitist multiobjective genetic algorithm: NSGA-II," in \textit{IEEE Trans. Evol. Comput.}, vol. 6, no. 2, pp. 182-197, Apr. 2002.

\bibitem{ref26} 
J. Párraga-Álava, M. Dorn and M. Inostroza-Ponta, ``Using local search strategies to improve the performance of NSGA-II for the Multi-Criteria Minimum Spanning Tree problem," 2017 IEEE Congress on Evolutionary Computation (CEC), Donostia, Spain, Jun. 2017, pp. 1119-1126.

\bibitem{ref27}
N. Krasnogor and J. Smith, ``A tutorial for competent memetic algorithms: model, taxonomy, and design issues," in \textit{IEEE Trans. Evol. Comput.}, vol. 9, no. 5, pp. 474-488, Oct. 2005.

\end{thebibliography}
\end{document}